\documentclass[a4paper,11pt]{article}
\pdfoutput=1
\usepackage[colorlinks=true]{hyperref}
\usepackage{graphicx,rotating,hyperref,slashed,amsmath,xcolor,amssymb,amsfonts,colortbl,cite}
\makeatletter
\hypersetup{colorlinks,bookmarksopen,bookmarksnumbered,  
linkcolor=blus,pdfstartview=FitH,urlcolor=rossos,citecolor=verde}
\allowdisplaybreaks		 

\hypersetup{%
    ,urlcolor=black
    ,citecolor=black
    ,linkcolor=black
    }
\usepackage{mciteplus}

\AtBeginDocument{
  \hypersetup{
    urlcolor=blue,
    citecolor=blue,
    linkcolor=black,
  }%
}

\def\edu{{\color{white}{]}}_{\large{|} \epsilon^{-2},\, \epsilon^{-1}}}
\def\ed{{\color{white}{]}}_{\large{|} \epsilon^{-2}}}
\def\eu{{\color{white}{]}}_{\large{|} \epsilon^{-1}}}
\def\bea{\begin{eqnarray}}
\def\eea{\end{eqnarray}}

\def\d{{\rm d}}

\def\d{{\rm d}}

\def\0{{\boldsymbol 0}}

\def\lsim{\mathrel{\rlap{\lower3pt\hbox{\hskip0pt$\sim$}}
   \raise1pt\hbox{$<$}}}         
\def\gsim{\mathrel{\rlap{\lower4pt\hbox{\hskip1pt$\sim$}}
   \raise1pt\hbox{$>$}}}         

 \newcommand{\sfootnote}[1]{}
\definecolor{bluc}{cmyk}{1,1,0,0.1}
\definecolor{rossoCP3}{cmyk}{0,.88,.77,.40}
\definecolor{rosso}{cmyk}{0,1,1,0.4}
\definecolor{rossos}{cmyk}{0,1,1,0.55}
\definecolor{rossoc}{cmyk}{0,1,1,0.2}
\definecolor{verdes}{cmyk}{0.92,0,0.59,0.4}

\hypersetup{colorlinks, bookmarksopen, bookmarksnumbered,
citecolor=verdes, linkcolor=bluc, pdfstartview=FitH, urlcolor=rossos}

\newcommand{\mio}[1]{}

\definecolor{Gray}{gray}{0.95}

\usepackage{multicol}
\usepackage{color}
\definecolor{rosso}{cmyk}{0,1,1,0.4}
\definecolor{rossos}{cmyk}{0,1,1,0.55}
\definecolor{rossoc}{cmyk}{0,1,1,0.2}
\definecolor{blu}{cmyk}{1,1,0,0.3}
\definecolor{blus}{cmyk}{1,1,0,0.6}
\definecolor{bluc}{cmyk}{1,1,0,0.1}
\definecolor{verde}{cmyk}{0.92,0,0.59,0.25}
\definecolor{verdec}{cmyk}{0.92,0,0.59,0.15}
\definecolor{verdes}{cmyk}{0.92,0,0.59,0.4}

\skip\@mpfootins = \skip\footins
\def\circa#1{\,\raise.3ex\hbox{$#1$\kern-.75em\lower1ex\hbox{$\sim$}}\,}

\newcommand{\be}{\begin{equation}}
\newcommand{\ee}{\end{equation}}
\newfam\rsfsfam
\def\mathscr#1{{\fam\rsfsfam\relax#1}}

\def\circa#1{\,\raise.3ex\hbox{$#1$\kern-.75em\lower1ex\hbox{$\sim$}}\,}
\makeatletter

\def\hhref#1{\href{http://arxiv.org/abs/#1}{arXiv:#1}} 

\newcommand{\doi}[1]{\href{http://dx.doi.org/#1}{[doi]}}

\def\hhref#1{\href{http://arxiv.org/abs/#1}{arXiv:#1}}

\def\art{\@ifnextchar[{\eart}{\oart}}
\def\eart[#1]#2#3#4#5#6{{\rm #2}, {\em #3 \bf #4} {\rm (#6) #5} ({\em #1})}

\def\article{\@ifnextchar[{\earticle}{\oarticle}}
\def\oarticle#1#2#3#4#5#6{{\rm #1}, {\em ``#6''}, {\rm #2 #3 (#5) #4}}
\def\earticle[#1]#2#3#4#5#6#7{{\rm #2}, {\em ``#7''}, {\rm #3 #4 (#6) #5}  [\hhref{#1}]}
\def\hepart[#1]#2{{\rm #2, \em#1}}
\def\heparticle[#1]#2#3{#2, {\em ``#3''} [\hhref{#1}]}

\newcounter{alphaequation}[equation]
\def\thealphaequation{\theequation\hbox to
0.6em{\hfil\alph{alphaequation}\hfil}}
\def\eqnsystem#1{
\def\@eqnnum{{\rm (\thealphaequation)}}
\def\@@eqncr{\let\@tempa\relax \ifcase\@eqcnt \def\@tempa{& & &} \or
  \def\@tempa{& &}\or \def\@tempa{&}\fi\@tempa
  \if@eqnsw\@eqnnum\refstepcounter{alphaequation}\fi
\global\@eqnswtrue\global\@eqcnt=0\cr}
\refstepcounter{equation} \let\@currentlabel\theequation \def\@tempb{#1}
\ifx\@tempb\empty\else\label{#1}\fi
\refstepcounter{alphaequation}
\let\@currentlabel\thealphaequation
\global\@eqnswtrue\global\@eqcnt=0 \tabskip\@centering\let\\=\@eqncr
$$\halign to \displaywidth\bgroup \@eqnsel\hskip\@centering
$\displaystyle\tabskip\z@{##}$&\global\@eqcnt\@ne
\hskip2\arraycolsep\hfil${##}$\hfil& \global\@eqcnt\tw@\hskip2\arraycolsep
$\displaystyle\tabskip\z@{##}$\hfil
\tabskip\@centering&\llap{##}\tabskip\z@\cr}
\def\endeqnsystem{\@@eqncr\egroup$$\global\@ignoretrue} \makeatother

\definecolor{fiorentina}{rgb}{.5,0,.5}

\begin{document}

\setcounter{page}{1} \baselineskip=15.5pt \thispagestyle{empty}

\bigskip\

\vspace{1cm}

\begin{center}
{ \selectfont \bf  \LARGE   Gravitational-wave cosmological distances\\ \,
 in scalar-tensor theories of gravity}
\end{center}

\vspace{0.2cm}

\begin{center}
{\fontsize{13}{30}\selectfont Gianmassimo Tasinato,$^{1}$  Alice Garoffolo,$^2$ Daniele Bertacca,$^{3,4}$ Sabino Matarrese $^{3,4,5,6}$}
\end{center}

\begin{center}
{\small
\vskip 8pt
\textsl{$^1$ Physics Department, Swansea University, SA28PP, UK}\\
\textsl{$^2$ Institute Lorentz, Leiden University, PO Box 9506, Leiden 2300 RA, The Netherlands}
\\
\textsl{$^3$
Dipartimento di Fisica e Astronomia Galileo Galilei, Universita' di Padova, I-35131 Padova, Italy}
\\
\textsl{$^4$
INFN Sezione di Padova, I-35131 Padova, Italy}
\\
\textsl{$^5$
INAF-Osservatorio Astronomico di Padova, Vicolo dell Osservatorio 5, I-35122 Padova, Italy}
\\
\textsl{$^6$
Gran Sasso Science Institute, Viale F. Crispi 7, I-67100 L'  Aquila, Italy}
\vskip 7pt
}
\end{center}

\smallskip
\begin{abstract}
\noindent
We analyze the propagation of high-frequency gravitational waves (GW) in scalar-tensor theories of gravity, with the aim of examining   properties of cosmological distances as inferred from GW measurements.   By  using symmetry principles, we first  determine the most general structure of the   GW linearized equations and of the GW energy momentum tensor,  assuming  that GW move with the speed of light.    Modified gravity effects are encoded in a small number of parameters, and we study the conditions for ensuring graviton number conservation in our covariant set-up. We  then apply our general findings to the case of   GW propagating   through a perturbed  cosmological space-time, deriving the expressions  for the GW luminosity distance  $d_L^{({\rm GW})}$ and  the GW angular distance $d_A^{({\rm GW})}$. We    prove for the first time the validity of Etherington reciprocity law  $d_L^{({\rm GW})}\,=\,(1+z)^2\,d_A^{({\rm GW})}$ for a perturbed universe   within a scalar-tensor framework.   We find that besides
 the GW  luminosity distance, also the GW angular distance can be modified with respect to General Relativity. We  discuss  implications  of this result for  gravitational lensing, focussing   on  time-delays of lensed GW and lensed photons emitted simultaneously   during   a multimessenger event.   We explicitly show how
 modified gravity effects  compensate between different   coefficients in the GW time-delay formula:   lensed GW   arrive at  the same  time as     their lensed electromagnetic  counterparts, in agreement with causality constraints.

\end{abstract}

\newpage

\section{Introduction}

The propagation of gravitational waves (GW) through cosmological distances
offer promising new avenues for testing cosmology. For
example, information on the GW luminosity distance as extracted standard siren events can be used to probe the distance-redshift
relation \cite{Schutz,Holz:2005df,Dalal:2006qt,MacLeod:2007jd,Nissanke:2009kt,Cutler:2009qv},
   leading to measurements of the present-day Hubble parameter
 using GW observations \cite{Abbott:2017xzu}. Also,
 gravitational wave measurements allow us  to test deviations from General Relativity (GR): in fact,
 GW results
 have been recently applied for excluding modified gravity models  predicting a speed of gravitational
 waves different than light \cite{Creminelli:2017sry,Sakstein:2017xjx,Ezquiaga:2017ekz,Baker:2017hug}, as earlier suggested in \cite{Lombriser:2015sxa,Bettoni:2016mij}. Measurements of the GW luminosity distance
 can also be used
  to probe modified gravity friction effects for GW travelling through cosmological backgrounds, see e.g. \cite{Belgacem:2017ihm,Deffayet:2007kf,Daniel:2008et,Saltas:2014dha,Sawicki:2016klv,Amendola:2017ovw,Nishizawa:2017nef,Arai:2017hxj,Belgacem:2018lbp,Lagos:2019kds,Mukherjee:2019wfw,Mukherjee:2019wcg,DAgostino:2019hvh,Vijaykumar:2020nzc,Mastrogiovanni:2020mvm,Mastrogiovanni:2020gua,Nishizawa:2019rra,Baker:2020apq,Mukherjee:2020tvr,Mukherjee:2020mha,Belgacem:2019tbw,Belgacem:2017cqo,Wolf:2019hun,MariaEzquiaga:2021lli}. In the future,
   precision GW measurements at high redshift, provided by LISA \cite{Audley:2017drz,Tamanini:2016zlh,Belgacem:2019pkk,Barausse:2020rsu} and the Einstein Telescope \cite{Maggiore:2019uih} will
    offer new possibilities for testing our understanding of gravity and cosmology.
   With  this aim in mind, it is imperative to further theoretically characterize the propagation of GW in alternative theories
    of gravity, also taking into account the implications  of cosmological inhomogeneities \cite{Laguna:2009re,Bertacca:2017vod,Fier:2021fbt} that might influence or be degenerate
    with modified gravity effects. This is the scope of this work,  concentrating on high-frequency scalar-tensor theories of gravity in the limit of geometric optics. We focus our analysis on propagation effects only,  assuming
    that at emission the properties of GW is identical to General Relativity.

    In section  \ref{sec-ourset} we  use of symmetry principles based on coordinate invariance
    for characterizing our scalar-tensor system and the behaviour of propagating degrees of freedom.
    We make use of a fully  covariant formulation, spelling
out in detail
 symmetry properties   under  coordinate transformations for
 each of the sectors  involved.
 This allows us to carry on a general, model
independent analysis of scalar-tensor systems, also identifying physically reasonable
conditions for decoupling the evolution  equations of different sectors.

Basing  our considerations  on symmetry principles based on coordinate
transformations, we then derive  in section  \ref{sec_theory2}
the most  general structure of the high-frequency GW evolution equations and energy-momentum tensor, for a  scalar-tensor set-up
in the limit of geometric optics.  We also define a covariant condition to express graviton number conservation in our
framework  (see also \cite{Belgacem:2018lbp}). Modified gravity effects factorize in front of our expressions, and the overall factor has a simple
physical explanation in terms of the modifications of the linearized evolution equations.

 We obtain  a set of covariant equations that can be used in a variety of situations. In section \ref{sec_pheno} we
apply them to the study of GW propagating through a perturbed cosmological space-time. In fact, distinguishing
implications of modified gravity from effects of cosmological perturbations will be a crucial step
for extracting physical information from future GW detections. Building and extending  the classic results by
Sasaki \cite{Sasaki:1987ad} (developed for studying propagation of photons in a perturbed cosmological universe
within General Relativity)
we
 derive the expressions  for the GW luminosity distance  $d_L^{({\rm GW})}$ and  the GW angular distance $d_A^{({\rm GW})}$. We    prove for the first time the validity of Etherington reciprocity law  $d_L^{({\rm GW})}\,=\,(1+z)^2\,d_A^{({\rm GW})}$ for a perturbed universe   within a scalar-tensor framework, for
  scenarios  where graviton number is conserved. Since this relation is at the basis for
relating angular and luminosity   distances in GW measurements, it is of crucial importance to
prove its validity in a general theory of gravity for GW propagation on a general space-time.
 (See  \cite{Arjona:2020axn} for a recent work discussing probes of Etherington reciprocity law using GW measurements.)
Given that GW luminosity distances can be modified with
respect to GR, also angular distances can receive corrections.

Values of  angular distances $d_A^{\rm (GW)}$ are important in phenomena
  involving strong lensing of GW, for example    for
   the time delay of lensed GW. Strong lensing of GW can be important  in the future
   for providing alternative ways for determining cosmological parameters (see e.g. \cite{Oguri:2019fix}).
When focussing on the limit of geometric optics for studying the propagation
   of GW and electromagnetic waves, since they both follow null-like geodesics we expect
   that GW and light arrive at the same time at the detector, if they are emitted at the same time \cite{Suyama:2020lbf,Ezquiaga:2020spg}. In section \ref{subsec-GWlensing} we explicitly
   show how to express the GW time-delay formula in terms of combinations of $d_A^{\rm (GW)}$, in such a way
   that all effects of modified gravity compensate  and one finds identical time-delays for GW and electromagnetic
   signals, if they are emitted simultaneously during a multimessenger event.

 Our conclusions can be found in section \ref{sec-conc}, and are followed by six technical appendixes.

\section{Our set-up}
\label{sec-ourset}

 We  develop  a covariant approach for investigating the dynamics of high-frequency modes in scalar-tensor
 theories of dark energy.  Symmetry arguments based on coordinate invariance   allow us to determine  general formulas describing the evolution
 of high frequency gravitational waves.

  Our  set-up is described by a covariant  action
 \be \label{genac1}
 S\,=\,\int d^4 x\,\sqrt{-g} \left[ \frac{M_{ P}^2}{2} R-{\cal L}(g_{\mu\nu}, \phi, matter)\right]\,,
 \ee
 coupling  gravity with a scalar field $\phi$ -- the dark energy (DE) field -- and  with additional matter
 fields,  schematically indicated with $matter$ in action \eqref{genac1}. We make the hypothesis that this action
 is invariant under   diffeomorphism transformations, i.e.  coordinate reparameterization  invariance:
 $x^\mu\to x^\mu+\xi^\mu(x)$ for arbitrary infinitesimal vector $\xi^\mu$.
  We do not need to further specify the  structure of the Lagrangian ${\cal L}$
 for our arguments, but in what follows we  assume that matter fields are minimally coupled
 with the metric $g_{\mu\nu}$ (possibly after performing  appropriate conformal transformations to select a Jordan frame). The dark energy field $\phi$, on the other hand, can have non-minimal
 kinetic couplings with the metric,
    that generally   influence the propagation of GW. 
     See e.g. \cite{Clifton:2011jh}
   for a comprehensive review on modified gravity models including scenarios with non-minimal couplings
   of scalars with the metric.

  One of the delicate issues  in studying GW propagation in modified gravity
    is to  distinguish tensor from scalar fluctuations, and correctly identify their roles in the evolution equations of high-frequency fields.  This topic  started with the classic papers
   \cite{Eardley:1973br,Eardley:1974nw}, and has been recently reconsidered
 in  \cite{Dalang:2019rke,Garoffolo:2019mna,Garoffolo:2020vtd,Dalang:2020eaj,Ezquiaga:2020dao} using a variety of methods.
  The issue can be subtle in theories where scalar and metric fluctuations propagate
  with different speed, a phenomenon  associated with spontaneous breaking of global Lorentz invariance
  by means  of a non-vanishing time-like  gradient for the dark energy field.
 Here we develop a  covariant approach to address the problem,  more
  similar in
 spirit to the original works of Isaacson and to the effective field theory of inflation \cite{Cheung:2007st} and dark energy \cite{Gubitosi:2012hu} (see e.g. \cite{Piazza:2013coa} for a comprehensive review). Our framework is
  distinct
    from    ones
      based on  decomposing  graviton helicities in terms of their rotational properties with respect to the GW axis of propagation.

      \subsubsection*{The perturbative expansion in high-frequency fields}

  We  base
  our considerations  on a double perturbative expansion for the metric and the scalar field around quantities solving the background
equations, as  \cite{Isaacson:1967zz,Isaacson:1968zza}. Schematically, we expand metric and scalar fields as
 \bea \label{dec1g}
g_{\mu\nu}(t, {\bf x})&=&\bar g_{\mu\nu}(t, {\bf x})+h_{\mu\nu}(t, {\bf x})\,,
\\ \label{dec1ph}
\phi(t, {\bf x})&=& \bar \phi(t, {\bf x})+\varphi(t, {\bf x})\,,
\eea
and we are interested to study the dynamics of the metric and scalar perturbations
 $h_{\mu\nu}$ and $\varphi$.
In the previous expression fluctuations
are distinguished from the background both for their
 absolute size   -- we call it {\it  expansion in the amplitude}, controlled by a  parameter $\alpha$ -- and for  the size of their gradients -- we call it {\it expansion in  gradients}, controlled by a parameter $\epsilon$. More specifically:
\begin{itemize}
\item[-]
The $\alpha$-expansion in the amplitude is
used  to define the so-called linear (first order) and quadratic (second order) approximations, and is
 common in cosmology. The parameter $\alpha$ is    a book-keeping
device to denote the order of
 amplitude expansion.
 \item[-]
The $\epsilon-$expansion in gradients is
 controlled by the physical quantity
 \be \label{defeps}
 \epsilon\,=\,\frac{\lambda}{L_B}\,,
 \ee
 controlling the ratio among the typical (small) wavelength $\lambda$
 of the high-frequency fields versus the  (large) scale $L_B$
 of spatial variation of slowly-varying background quantities. Among
 the latter, we include a dark energy scalar  $\bar \phi(x)$ whose
 time-like profile varies on scales of order $L_B$.
 \end{itemize}
 The fluctuations $h_{\mu\nu}$ and $\varphi$ are thought as high-frequency fluctuations whose gradients
 are enhanced by a factor of $1/\epsilon$ with respect to the background; moreover, they are small perturbations
 whose amplitude is suppressed by a factor of order ${\cal O}(\alpha)$ with respect to the background.

 The possibility to use  $\epsilon$  as small parameter to organize a perturbative expansion is one of the
 key observations of Isaacson: his approach is  reviewed and expanded in the
 textbooks  \cite{Misner:1974qy,Maggiore:1900zz}.
  We adopt it here, extending the discussion of  \cite{Garoffolo:2019mna}.
  This framework  allow us to implement  a geometric optics limit where a generic
small fluctuation $\sigma(x)$ (scalar or metric) is decomposed
 into a slowly-varying amplitude, and a rapidly-varying
phase
 (we  understand the `real part' symbol in what follows)
\be
\label{defsig1a}
\sigma(x)\,=\,{\cal A}_\sigma(x)\,\exp{\left(\frac{i \psi_\sigma(x)}{\epsilon}\right)}\,.
\,
\ee
 $\epsilon$ is the small parameter of eq \eqref{defeps} controlling the rapid phase variations. The evolution equations
contain up to second order derivatives in the fields: hence, substituting Ansatz
\eqref{defsig1a} in such equations, we expect contributions scaling as $1/\epsilon^2$, $1/\epsilon$,
 plus positive (or null) powers of $\epsilon$. The geometric optics framework
 focusses on the leading ($1/\epsilon^2$) and next-to-leading ($1/\epsilon$)
 orders in expansion in the small parameter $\epsilon$ -- controlling respectively
 the evolution of phase and amplitude -- and neglects the higher-order terms.

      \subsubsection*{The symmetry transformations}

We consider  coordinate transformations acting on the quantities
$h_{\mu\nu}$ and $\varphi$.
 We denote the gradient of the
  low-frequency scalar mode  profile as
\be
\label{defgrad1}
 \nabla_\mu\, \bar \phi\,=\,v_\mu
 \,,
\ee
 spontaneously breaking  global coordinate reparametrization along the direction of the time-like vector $v^\mu$.
 The vector $v^\mu$ can be thought as being associated with cosmological acceleration, analogously to the approaches
 of the effective field theory for inflation and dark energy \cite{Cheung:2007st,Gubitosi:2012hu}, and
  plays a special role in our discussion.
    From now on, all covariant derivatives are taken with respect
 to the low-frequency metric field $\bar g_{\mu\nu}$ of eq \eqref{dec1g},  used also to raise and lower indexes.
The non-vanishing gradient \eqref{defgrad1}
 has important implications for diffeomorphism transformations. Under a change of coordinates the
 linearized fluctuations transform as
 \bea
\label{trasf1}
h_{\mu\nu}&\to&h_{\mu\nu}- \nabla_\mu \xi_\nu- \nabla_\nu \xi_\mu\,,
\\
\label{trasf2}
\varphi&\to&\varphi- v^\mu \,\xi_\mu\,,
\eea
for infinitesimal vector $\xi_\mu$.
  The scalar symmetry transformation \eqref{trasf2} corresponds to  a non-linearly realized diffeomorphism
 transformation, after the spontaneous space-time symmetry breaking associated with the
 scalar gradient $v^\mu$.

 We assume from now on that the amplitude of $h_{\mu\nu}$ is of order ${\cal O}(\epsilon^0)$
 in a gradient expansion, and we neglect in what follows possible contributions of order  ${\cal O}(\epsilon^1)$
 and higher in the $\epsilon$-parameter.  (We checked that, even including those contributions, the arguments
 we develop are all still valid.)
In order to actively apply the transformation on the fast moving modes $h_{\mu\nu}$ we assume that
the size of $\xi_\mu$  is reduced by a factor of $\epsilon$ with respect to $h_{\mu\nu}$. i.e.
\be
\label{ampscal1}
{\cal O}\left(
\xi_\mu
\right)
\,\sim \, \epsilon \,
{\cal O}\left(
h_{\mu\nu}
\right)\,.
\ee
The gradients of $\xi_\mu$ in eq \eqref{trasf1}
enhance its contributions by a factor ${\cal O}(1/\epsilon)$, so that the result is of order
 ${\cal O}(1/\epsilon) \times {\cal O}(\epsilon)\,=\, {\cal O}(\epsilon^0)$, i.e.
  of the same order of $h_{\mu\nu}$ in an $\epsilon$-expansion. Again,  for simplicity  we assume that $\nabla_\mu \xi_\nu$
does not receive  contaminations at order ${\cal O}(\epsilon^1)$,
since as mentioned above we neglect contributions of order ${\cal O}(\epsilon^1)$ and higher
to the metric fluctuations $h_{\mu\nu}$.

\smallskip
What can we say about the size of $\varphi$? We start noticing that the symmetry transformation
\eqref{trasf2} `turns on' high-frequency scalar excitations even if they are initially absent.
The  dynamics of the two sectors, metric and DE perturbations,  is inevitably coupled in their path from emission to detection.  Even if DE fluctuations are not produced at the source (for
example thanks to some screening mechanism), they can be generated by metric fluctuations that are travelling from
source to detection.
We then expect that propagation effects are able to excite
  scalar modes
 with an  amplitude
 suppressed by a factor of ${\cal O}(\epsilon)$
with respect to metric fluctuations:
\be\label{scalhie1}
{\cal O}(\varphi)\sim \epsilon\, {\cal O}( h_{\mu\nu})\,.
\ee
Then,  scalar modes transform non-trivially under the non-linearly realized diffeomorphism
transformations controlled by the quantity $v^\mu\,\xi_\mu$ (which is of order $ {\cal O}(\epsilon^1)$).

\smallskip
Our expectation encoded in the hierarchy \eqref{scalhie1} is also supported by interpreting
  scalar fluctuations $\varphi$ as `Goldstone bosons' of  global space-time symmetries
 broken by the scalar profile $v^\mu$ \cite{Cheung:2007st,Gubitosi:2012hu}.  The size of
  the background
gradient is of order  $\nabla_\mu \bar \phi\,\sim \,L^{-1}_B$.
 Keeping a fixed high-frequency wavelength $\lambda$ for the metric fluctuations $h_{\mu\nu}$, in the limit $\nabla_\mu  \bar \phi \to0$ (or equivalently $L_B\to \infty$)  we expect the scalar Goldstone modes $\varphi$ to be absent, since the symmetry is restored, and Goldstone bosons do not propagate.  The size of the scalar excitation $\varphi$ can be  then expected to be suppressed by a factor
$\lambda/L_B\sim \epsilon$ with respect to  metric fluctuations, in agreement with eq \eqref{scalhie1}.
    Motivated by these arguments, we impose
 the hierarchy \eqref{scalhie1} for linearized fluctuations.

\smallskip

Under these hypothesis, in the technical appendix \ref{sec_theory1} we build
combinations of scalar and metric fluctuations that transform conveniently under
coordinate transformations.  After appropriate gauge fixings, we single out a transverse-traceless
dynamical fluctuation $h_{\mu\nu}^{(TT)}$ of order ${\cal O}(\epsilon^0)$, orthogonal to the vector $v^\mu$,  that we identify with the  high-frequency GW. Its dynamics is invariant
under the residual transformation
\be\label{finressA}
h_{\mu\nu}^{(TT)}\,\to\,h_{\mu\nu}^{(TT)}-\nabla_\mu \xi_\nu^{(T)} -\nabla_\nu \xi_\mu^{(T)}\,,
\ee
where $\xi_\mu^{(T)}$ is a vector orthogonal to  $v^\mu$, which satisfies additional  gauge conditions we spell out in appendix
 \ref{sec_theory1}. The remaining high-frequency degrees of freedom are scalar modes. 
We develop arguments to show that, under the condition that  scalar and GW propagate with different velocities
 along different geodesics, such
   scalar perturbations decouple from GW modes at linear order in perturbations.
 From now on, for definiteness, we then concentrate on studying the dynamics of the transverse-traceless
 GW modes $h_{\mu\nu}^{(TT)}$, leaving the study of the independent scalar sector, when propagating~\footnote{Proposals exist for building scalar-tensor theories that do not propagate scalar modes -- see e.g. \cite{Tasinato:2020fni}.}, to a separate
 work.

\section{GW evolution  equations and energy momentum tensor}
  \label{sec_theory2}

 Symmetry considerations
provide a powerful tool for constraining the dynamics
 of our system.  In fact, we can use symmetry arguments to determine
 the structure of the linearized GW equations of motion and energy-momentum
 tensor, with no need to rely on specific models. This is the aim of this section.


\subsection{The linearized evolution equations}\label{sec-lin}

 Isaacson, working in the context of the geometric optics limit of  General Relativity (GR), shown that the original diffeomorphism
 invariance  is preserved order-by-order in the gradient expansion,
 and at each order in $\epsilon$ the system is invariant under coordinate transformations \cite{Isaacson:1967zz,Isaacson:1968zza}.
This property further demonstrates  the utility of the perturbative scheme based on gradients, which
can be made compatible with the symmetries of the original theory, at least within the limits
of geometric optics. We now make  use of this
fact in the scalar-tensor framework we are interested in. We change perspective
and {\it impose} the symmetry invariance of the evolution equations
at each order in the $\epsilon$-expansion. As we will see,  this viewpoint allows us to write the most
general structure for the equations governing the GW dynamics, and to encode
the effects of modified gravity in few physically transparent parameters.

Our starting point are the  Einstein equations
for high-frequency GW fluctuations, expanded at first order ${\cal O}(\alpha^1)$ in
the amplitude.
Calling $G_{\mu\nu}$ the Einstein tensor, and  with the  suffix $(n)$ the order of the amplitude
$\alpha$-expansion, the system
of equations can be written as
\bea
\label{linEQ1}
G_{\mu\nu}^{(1)}\left[ h^{(TT)}_{\rho\sigma} \right]\edu&=&T^{(1)}_{\mu\nu}\left[ h^{(TT)}_{\rho\sigma} \right] \edu \hskip1cm,\,\, {\text{at order ${\cal O}(\alpha^1)$\,,}}
\eea
where $ h^{(TT)}_{\mu\nu } $ are the transverse-traceless fluctuations, orthogonal to the vector $v^\mu$,  that we identify with GW -- see the discussion
around eq \eqref{finressA}.
 We focus our attention to the leading and next-to-leading orders
 $\epsilon^{-2}$ and   $\epsilon^{-1}$ in the $\epsilon$ gradient expansion, which
 define the geometric optics framework as discussed after eq \eqref{defsig1a}.  Such contributions
 are obtained by singling out terms containing respectively second and first derivatives on the fields involved.
 As stated at the end
 of the previous section, we focus on the evolution of the GW tensor $ h^{(TT)}_{\mu\nu } $ only, and
 postpone an analysis of dynamically independent high-frequency scalar modes  to a separate work.

  In the usual geometric optics Ansatz of GW propagation in GR, it is costumary to assume that matter
 fields are slowly-varying, and one considers
 the evolution equations at order $1/\epsilon^2$ and $1/\epsilon$ as free equations ($T_{\mu\nu}=0$). %
  Here we
  go beyond this hypothesis.  In fact, a dark energy  scalar field
   plays an important role in determining the behaviour of gravity at large cosmological distances.
     Kinetic
   couplings between scalar and metric lead to derivatives acting on the high-frequency modes, contributing
    to the effective linearized EMT $T^{(1)}_{\mu\nu}$.  Non-minimal couplings
    between the dark-energy scalar and the metric are in fact common and well-motivated in theories
    of dark energy and modified gravity.

Nevertheless,
 symmetry considerations allow us to determine the general structure of $T^{(1)}_{\mu\nu}$, without relying on specific models.
We demand that GW propagate
with the speed of light. Since $h_{\mu\nu}^{(TT)}$ satisfies a
  transverse-traceless gauge, as well as the orthogonality requirement
  $
v^\rho
 h^{(TT)}_{\rho\sigma}\,=\,0$,
one finds that the left-hand-side of \eqref{linEQ1} reads
\be
\label{newoldco1c}
G_{\mu\nu}^{(1)} \left[ h^{(TT)}_{\rho\sigma} \right]\edu\,=\,-\frac12\,
\Box  h_{\mu\nu}^{(TT)}
\edu\\.
\ee
We can now discuss the allowed structure for
 the linearized energy-momentum tensor $T^{(T)}_{\mu\nu}$ contributing  to the GW evolution equation at
  orders $\epsilon^{-2}$ and $\epsilon^{-1}$. It should be
   transverse-traceless, and orthogonal to $v^\mu$ at orders $\epsilon^{-2}$ and $\epsilon^{-1}$; moreover,
   it should be
  conserved at order $\epsilon^{-2}$: $\left[\nabla^\mu\,T^{(T)}_{\mu\nu} \right]_{{\epsilon^{-2}}}\,=\,0$,
  and it should be invariant under the transformation of eq \eqref{finressA}.
   Finally, we demand that it ensures that GW propagate with light speed, to be consistent with GW170817 constraints
   \cite{Abbott:2018lct},  and have then
  a standard dispersion relation.

  The only allowed structure of the linearized  $T^{(T)}_{\mu\nu} ( h_{\rho\sigma})$ that satisfies these requirements
at orders $\epsilon^{-2}$, $\epsilon^{-1}$
is
\bea
T^{(T)}_{\mu\nu}&=&\tau_A \,\Box  h_{\mu\nu}^{(TT)}
+\tau_B
\,
v^\rho\,\nabla_{\rho}  h^{(TT)}_{\mu\nu}
  \label{TMUNUstr}\,,
\eea
where $\tau_{A, B}$ depend only on slowly varying fields.
 In fact, at order $\epsilon^{-2}$, $T^{(T)}_{\mu\nu}$ contains second derivatives, but the  unit-speed condition
 only allows for the combination proportional to $\tau_A$ in the formula above. At order $\epsilon^{-1}$ it contains first derivatives, but the gauge  conditions we impose allow only  for the contribution proportional to  $\tau_B$
 in eq \eqref{TMUNUstr}.
 Calling  the combination
 \be
 \label{defTg}
 {\cal T}\,=\,-\frac{2\,\tau_B}{1+2\tau_A}\,,
 \ee
  which depends on slowly-varying fields only,
 we  rewrite the linearized evolution equation for GW fluctuations in terms of a single parameter characterizing deviations
 from GR:
 \be \label{singpa1}
\left( \Box  h_{\mu\nu}^{(TT)} \right) \edu\,=\,{\cal T}\,\times \left( v^\rho \nabla_\rho\, h_{\mu\nu}^{(TT)} \right)\eu \,.
 \ee
 The deviations from GR on the propagation of high-frequency GW only appear as a first-order gradient of the GW high-frequency fluctuation, proportional
 to the parameter ${\cal T}$ depending on slowly-varying fields. Such contribution can be thought as a `friction term' for the GW, and is common
 to find it in scalar-tensor systems with non-minimal kinetic couplings between scalar and metric degrees
 of freedom.  In the context of gravitational wave cosmology  in modified gravity several groups   explored
  the consequences of such friction term in specific  cosmological models, see e.g.
  \cite{Belgacem:2017ihm,Daniel:2008et,Saltas:2014dha,Sawicki:2016klv,Amendola:2017ovw,Nishizawa:2017nef,Arai:2017hxj,Belgacem:2018lbp,Lagos:2019kds,Mukherjee:2019wfw,Mukherjee:2019wcg,Vijaykumar:2020nzc,Mastrogiovanni:2020mvm,Mastrogiovanni:2020gua,Nishizawa:2019rra,Baker:2020apq}.
  (see also \cite{Ezquiaga:2018btd} for a review), finding it is related with the parameter called $\alpha_M$ in the effective field theory approach to dark energy.
   Also, this friction term arises
  in cosmological models with time-varying Planck mass: see Appendix \ref{exafphr} for the analysis of a representative model~\footnote{It would be interesting to extend our arguments to set-up with space-time
  dimensions different than four \cite{Pardo:2018ipy,Calcagni:2019kzo,Calcagni:2019ngc}, which can not be directly expressed in terms of a covariant four dimensional
  action.}.
   It is interesting to find that the fully covariant  `beyond-GR' friction term in eq \eqref{singpa1}  is the only one allowed  by our symmetry
  principles and our physical considerations. In what follows, we consider the quantity ${\cal T}$
  as an effective  parameter controlling deviations from General Relativity.

\subsection{Evolution equations in the limit of geometric optics }\label{secGeoAN}

We now discuss how our covariant  equations at leading and next-to-leading orders in an $\epsilon$-gradient expansion
allow us to derive the evolution equations for the physical degrees of freedom in the limit of geometric
optics. As stated above, we focus only on the GW sector controlled by the transverse-traceless tensor
$h_{\mu\nu}^{(T)}$.

 The eikonal Ansatz for the GW reads
\bea
 h_{\mu\nu}^{(TT)}&=&{\cal A}_T\,{\bf e}_{\mu\nu}\,\exp{\left[
i\,{\psi^{(T)}}/{\epsilon}
\right]
}
\label{defAnsh}\,,
\eea
with ${\cal A}_T$ the amplitude, $\psi^{(T)}$ the phase, $\epsilon$ the small parameter of eq \eqref{defeps}, and
${\bf e}_{\mu\nu}$ a polarization tensor normalized such that ${\bf e}_{\mu\nu}\,{\bf e}^{\mu\nu}\,=\,1$.
The gradient of the phase defines the GW 4-momentum\footnote{We choose   the same conventions of \cite{Sasaki:1987ad} for the overall sign in the definition of the wave vector.}:
\be
k_\mu\,=\,\,\nabla_\mu \psi^{(T)}\,.
\ee
We now apply Ansatz \eqref{defAnsh}  to the covariant evolution equation  \eqref{singpa1}, and separate
the geometric optics analysis of the orders $1/\epsilon^2$ and $1/\epsilon$ in our gradient expansion. The transverse-traceless condition, and the condition
of orthogonality with respect to $v^\mu$ impose the following requirements on the polarization tensor:
\be
{\bf e}_{\mu}^{\,\,\mu}\,=\,k^\mu\,{\bf e}_{\mu\nu}\,=\,v^\mu \,{\bf e}_{\mu\nu}
\,=\,0\,,
\ee
where notice that there is a degeneracy among the last two conditions, so the previous equations
provide 8 instead of 9 independent conditions.
The order $1/\epsilon^2$ of the equations, obtained
from singling out second derivatives on the fields,  control the evolution of the GW phase and the GW
dispersion relations, leading to
\be\label{con1ng}
k^\mu k_\mu\,=\,
k^\rho \nabla_\rho k^\mu
\,=\,
0\,.
\ee
I.e. the GW 4-momentum is a null vector, propagating along a null geodesics. It is convenient to define
the affine parameter $\lambda$ controlling the evolution along the GW geodesics: for any function $f$,
the derivative along the affine parameter is defined by
\be
\frac{d f}{d \lambda}\,\equiv\,k^\rho \nabla_\rho f\,.
\ee
The integral curves of the vectors $k^\mu$ define the GW-rays:
\be
\frac{d x^\mu}{d\lambda}\,=\,k^\mu\,,
\ee
an important quantity for what follows.

\smallskip

While so far nothing changes with respect to General Relativity,
at order $1/\epsilon$  -- obtained from first derivative contributions -- the effects of modified gravity become manifest. The evolution equation for
the amplitude is
\be
\label{evowT}
\left[
2 \,k^\rho \nabla_\rho  \,{\cal A}_T+ (\nabla_\rho k^\rho)  \,{\cal A}_T
\right]\,=\,{\cal T}\,
 k^\rho v_\rho\,
{\cal A}_T\,,
\ee
where the quantity ${\cal T}$, given in eq \eqref{defTg}, depends on slowly-varying fields only. Recalling that
$v_\mu\,=\,\nabla_\mu \bar \phi$, the previous equation
 can be `integrated' to
\be\label{evamp1}
\nabla_\rho \left( e^{-\int {\cal T}}\,k^\rho {\cal A}_T^{2} \right)\,=\,
0\,.
\ee

The schematic expression  $\int {\cal T}$ denotes the following integral
\be \label{defoI}
\int {\cal T}\,\equiv\,\int_{\lambda_s}^\lambda {\cal T} \,\frac{d  \bar \phi}{d  \lambda'}\,d \lambda'\,,
\ee
with $\lambda_s$ corresponding to the value of the affine parameter at the source position.
The   quantity \eqref{defoI} represents
 a cumulative integration
 of modified gravity effects (the friction term in eq \eqref{singpa1}) over
 the
 the GW geodesic's affine parameter.   In integrating eq \eqref{evowT} we have chosen boundary conditions so that modified
  gravity contributions vanish at the location $\lambda\,=\,\lambda_s$ of the source, as expected
  since near emission modified propagation effects do not have time to develop. Modified gravity effects
  get exponentiated and appear as an overall factor inside the parenthesis in  equation \eqref{evamp1}: importantly, we do not need to demand that ${\cal T} $ is `small' for writing the equation. The exponential structure above
  will have several interesting consequences for our discussion.

\subsection{The energy momentum of GW at second order in perturbations}\label{sec-GWEMT}

 Isaacson \cite{Isaacson:1968zza} proved that GW can be associated with  their own energy-momentum-tensor  (EMT), defined at second
 order in the $\alpha$-expansion, which can influence the background dynamics. Schematically, we can write
 \bea
 \label{emt2o}
G_{\mu\nu}^{(0)}&=&\epsilon^2\,T^{(2)}_{\mu\nu}\ed  \hskip1cm,\,\, {\text{up to order ${\cal O}(\alpha^2)$}}\,,
\eea
where $T^{(2)}_{\mu\nu}$ denotes the  EMT associated with GW. As stated above, we focus
on the contributions associated with the  transverse-traceless tensor fluctuations  $ h^{(TT)}_{\mu\nu}$ only,
and do not discuss scalar contributions in this work since, under our hypothesis, the two sectors
evolve independently.
 The quadratic terms in  the GW energy-momentum-tensor have equal-size momenta in  opposite
directions which compensate each other, hence contributing at zeroth order in the $\epsilon$-expansion.
Using only symmetry arguments  we are able to determine the structure of the GW contribution
to the tensor $T^{(2)}_{\mu\nu}$ in a general class of scalar-tensor systems.

    When focussing on transverse-traceless excitations,
   Isaacson's result for the EMT is

\be\label{tmngw1}
T_{\mu\nu}^{\rm (2),\,GR}\,=\,
\epsilon^{2}\,\frac{{1}}{32\,\pi}\,
\,\langle \nabla_{\mu}  h^{(TT)}_{\rho\sigma}
\,\nabla_{\nu}  h^{(TT),\,\,\rho\sigma}\rangle\,.
\ee
 The symbol $\langle \dots \rangle$
denotes the so-called
 Brill-Hartle spatial average, see  \cite{Isaacson:1968zza,Misner:1974qy}. Among other things,  this average
 procedure ensures that the EMT is
 diffeomorphism-invariant, and conserved.

 Interestingly, the condition of coordinate invariance   fixes the structure of
 $T_{\mu\nu}^{\rm (2)}$ associated with GW.
 In fact,
 recall the EMT is quadratic in $ h^{(TT)}_{\mu\nu}$, and contains
two derivatives in total acting on the transverse-traceless GW excitations (by `integration by parts', we can place one derivative per field).
The structure in the combination \eqref{tmngw1} within the average is
the only one with these properties, and that is compatible with the condition of invariance under symmetry \eqref{finressA}. See the discussion in Appendix \ref{appA}.

The only freedom we are left with is in the overall factor in front of the Brill-Hartle  average appearing in eq \eqref{tmngw1}.
In fact we can change perspective, and use the condition of  invariance under symmetry for
determining the structure of  $T_{\mu\nu}^{\rm (2)}$
in  the scalar-tensor set-up we are interested in. In other words,
we do not compute the EMT using a `top-down' approach
starting from a given theory, but  instead we deduce its structure from the symmetry  conditions imposed in the theory.
In the scalar-tensor framework we are focussing on, the previous considerations
 allow for the following structure for the EMT in modified gravity
\be\label{tmngw2}
T_{\mu\nu}^{\rm (2),\,ST}\,=\,\epsilon^{2}\,\frac{\#}{32\,\pi}\,\langle \nabla_{\mu}  h^{(TT)}_{\rho\sigma}
\,\nabla_{\nu}  h^{(TT)\,\,\rho\sigma}\rangle\,,
\ee
where $\#$ is a function (to be determined) of the slowly-varying fields, metric and scalar. We now proceed to determine this quantity, making
use of the condition that the energy-momentum tensor $T_{\mu\nu}^{\rm (2),\,MG}$ should be conserved
by virtue of the Bianchi identity, and of the geometric optics evolution equations of section \ref{secGeoAN}.
  We substitute
the geometric Ansatz of section \ref{secGeoAN} to the previous formula, and get
\be\label{tmngw2a}
T_{\mu\nu}^{\rm (2),\,ST}\,=\,\frac{\#}{32\,\pi}\,{\cal A}_T^2\,k_\mu k_\nu\,.
\ee
Using the evolution equation \eqref{evamp1}, as well as the condition \eqref{con1ng} that GW follow null-like
geodesics, the
 condition of conservation of  the EMT
 $$
\nabla^\mu T_{\mu\nu}^{\rm (2),\,ST}\,=\,0\,,
$$
fixes $\#$ to the value ${e^{-\int {\cal T}}}$ as defined~\footnote{As commented
after eq \eqref{defoI}, the boundary conditions on the integral are chosen
such to ensure that at the position of the source the effects
of modified gravity vanish.} in eqs \eqref{evamp1}, \eqref{defoI}:
\be \label{defoIA}
e^{-\int {\cal T}}\,=\,e^{-\int_{\lambda_s}^\lambda {\cal T} \,\frac{d  \bar \phi}{d  \lambda'}\,d \lambda'\,}.
\ee
 Hence we find that the second order GW energy-momentum tensor in our scalar-tensor
 framework, in the geometric optics limit, reads
\be\label{tmngwb}
T_{\mu\nu}^{\rm (2),\,ST}\,=\,\frac{e^{-\int {\cal T}}}{32\,\pi}\,{\cal A}_T^2\,k_\mu k_\nu \,.
\ee
These general considerations then allow us to single out  transparently the effects of modified gravity
in the overall factor depending on the quantity $\int {\cal T}$ of eq \eqref{defoI}, a cumulative
integral of modified gravity contributions along the GW geodesics from source to detection. As we will learn in what  follows,
 phenomenological implications of our results, as well as the explicit example discussed in section \ref{exafphr}, further support the structure \eqref{tmngwb}
for the EMT in the scalar-tensor systems under consideration.

\subsection{Conservation of graviton number}\label{sec-consgn}

We can do some further steps following \cite{schneider}, and relate the properties of the quantities above
with graviton number conservation.  we expect
 graviton number to be conserved within a GW ray bundle, and we are going to prove 
 this fact in our setting within geometric optics. 
 We express $T_{\mu\nu}^{\rm (2),\,ST}$ in terms of quantities $k^\mu$ -- interpreted as graviton 4-momentum -- and
${\cal N}^\mu$, as:
\be
T^{\rm (2),\,MG}_{\mu\nu}\,=\,\frac{1}{32\,\pi}\,k_\nu \, {\cal N}_\mu\,,
\ee
where $ {\cal N}_\mu$ is interpreted as the graviton number density, and is defined as
\be\label{granuc}
 {\cal N}_\mu\,\equiv \,k_\mu\,{\cal A}_T^2\,e^{-\int {\cal T}}\hskip1cm\Rightarrow\hskip1cm \nabla_\mu
 \, {\cal N}^\mu\,=\,0\,.
\ee
Graviton-number conservation $\nabla_\mu
 \, {\cal N}^\mu\,=\,0$ is ensured by
relation \eqref{evamp1} \footnote{Graviton number conservation is a consequence of the fact that GW and scalar excitations are decoupled in our framework, since they travel along different geodesics in the limit of geometric optics. It would be interesting to understand the corresponding conditions in scenarios with direct couplings
among the two sectors, in models where scalar and GW high-frequency modes
move with the same speed \cite{Garoffolo:2019mna,Garoffolo:2020vtd,Dalang:2020eaj,Ezquiaga:2020dao}.}.
 See also \cite{Belgacem:2017ihm} for a  perspective on
 graviton number conservation in a cosmological setting in a modified gravity framework.

\smallskip

Moreover,
 by making use of well-known
geometric relations (see Figure \ref{fig:cons}),
 we can express this condition in a geometrically more direct way, which further supports our
  identification of ${\cal N}^\mu$ with graviton number density.
 \begin{figure}
\centering
 \includegraphics[width = 0.42 \textwidth]{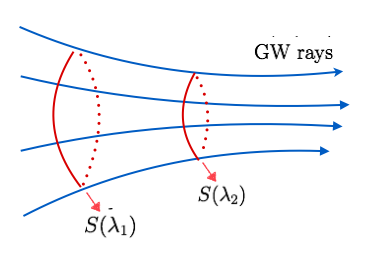}
 \caption{\it Geometric optics representation of graviton number conservation.
 The flux
 of
  a stream of gravitons crossing the  $S$-areas is conserved along the GW affine parameter. See eq \eqref{evSAv2}.  }
 \label{fig:cons}
\end{figure}
We call $S(\lambda)$ the cross-sectional area of a GW bundle, and $\lambda$ the
affine parameter along each GW ray (the graviton trajectory) with four-momentum $k^\mu$. A geometric
optics theorem (see \cite{Misner:1974qy}, exercise 22.13) states that
\be \label{evSv1}
\frac{d\,S(\lambda)}{d \lambda}-\nabla_\mu k^\mu \,S(\lambda)\,=\,0\,,
\ee
with $\lambda$ the affine parameter associated with the GW 4-momentum $k^\mu$.
Together with  \eqref{granuc}, relation \eqref{evSv1} implies the important identity

\be  \label{evSAv2}
\frac{d}{d \lambda}
\left\{
 e^{-\int {\cal T}}\,
 {\cal A}_T^2\, S
 \right\}
\,=\,0\,,
\ee
that  makes  more manifest the required flux conservation for a stream of gravitons crossing the  $S$-areas along the
GW evolution parameterized with $\lambda$.  Notice the  presence   of the overall exponential factor $e^{-\int {\cal T}}$
due to modified gravity -- see the discussion after eq \eqref{defoI} -- that changes
the flux of gravitons through a given surface. Such coefficient plays the role of   `damping term' in the GW amplitude during propagation in a modified gravity set-up,
as expected given that we interpret ${\cal T}$ as friction in the evolution equations.
 The result \eqref{evSAv2}
will be important for  cosmological applications
in what follows.

\section{Cosmological  distances and GWs}
  \label{sec_pheno}

  We now apply the general findings of the previous sections to GW propagating through
  a perturbed Friedmann-Robertson-Walker (FRW)  space-time.
   We  prove the validity of Etherington reciprocity law between GW luminosity and angular distances
  in the  scalar-tensor framework developed in the previous sections, and
we  discuss the implications of our findings for GW lensing.

    \smallskip

  Cosmologists use
  various different definitions of {\it distance} depending on the context,  and the
  observables they are interested in (see e.g.  \cite{Hogg:1999ad,Chen:2017wpg} for  enlightening reviews). While    usually definitions make use
   of light detected from distant sources, GW offer  new tools for measuring
  cosmological distances. We consider here two distinct  GW distance probes:
  \begin{enumerate}
  \item The {\it GW luminosity distance $d_L^{\rm ( GW)}$} is defined in terms of the ratio of GW power emitted at source position (intrinsic GW luminosity), versus the GW flux at detector location -- see section \ref{sec-dL}.
  The luminosity distance
  depends on the universe expansion rate, and
   enters into the GW waveforms and can be directly measurable by detecting GW  from
  distant sources. Following early important works \cite{Schutz,Holz:2005df,Dalal:2006qt,MacLeod:2007jd,Nissanke:2009kt,Cutler:2009qv},  $d_L^{\rm (GW)}$ is being
  recognized as a key
  observable to independently  measure cosmological parameters
 by means of  GW, as well
  as testing theories of modified gravity (see e.g. the review \cite{Ezquiaga:2018btd}).
    \item { The {\it GW angular distance $d_A^{\rm (GW)}$} is formally defined in terms of the ratio between
    the source angular diameter 
         at emission, versus the source angular size
    at detector location -- see section \ref{sec-dA}.
       Presently the angular resolution of GW detectors is not very
    accurate, although in the future it can increase, if
         more sophisticated  instruments  become available (see e.g. \cite{Baker:2019ync}). The quantity $d_A^{\rm (GW)}$ is  important in the context of {\it GW lensing}, a subject  with interesting possibilities for GW physics -- see e.g.  \cite{Turner:1990mk,Wang:1996as,Nakamura:1997sw,Wang:1999me,Macquart:2004sh,Takahashi:2003ix,Seto:2003iw,Yoo:2006gn,Seto:2009bf,Shapiro:2009sr,Sereno:2010dr,VanDenBroeck:2010fp,
Sereno:2011ty,Baker:2016reh,Kyutoku:2016zxn,Dai:2017huk,Oguri:2018muv,Dai:2018enj,Haris:2018vmn,Oguri:2018muv,Hou:2019dcm,Liu:2019dds,Meena:2019ate,Hannuksela:2019kle,Morita:2019sau,
Cusin:2020ezb,Pang:2020qow,Rubin:2020gba,Ezquiaga:2020dao,Ezquiaga:2020gdt,Ezquiaga:2020spg,Takahashi:2016jom,Suyama:2020lbf,Contigiani:2020yyc}
  for papers discussing the topic from a variety of perspectives.}
  \end{enumerate}

The angular diameter distance  $d_A$ is usually understood as being related with the luminosity distance
$d_L$ through the so-called duality-distance relation, or Etherington reciprocity law $d_L\,=\,(1+z)^2\,d_A$. On
the other hand, the theoretical validity of this relation should be explicitly proved, and this is of the  aims of this section, together with applications to GW lensing.
 In fact,
   since we learned in the previous section that GW evolution is
  affected by the friction term proportional to ${\cal T}$ in eq \eqref{evowT}, we expect that both luminosity and
  angular distances are influenced by modified gravity. We show that,
 thanks to graviton number conservation (see section \ref{sec_theory2}),
 these quantities are related
 by
     Etherington reciprocity law
       (see e.g. \cite{schneider,Etherington}  for the case of photon propagation)
  \be\label{GWether}
  d_L^{\rm (GW)}\,=\,(1+z)^2\,d_A^{\rm (GW)}\,,
  \ee
  for GW propagating through perturbed FRW space-times in the scalar-tensor scenarios
  we are focussing on.

 \begin{figure}
\centering
 \includegraphics[width = 0.42 \textwidth]{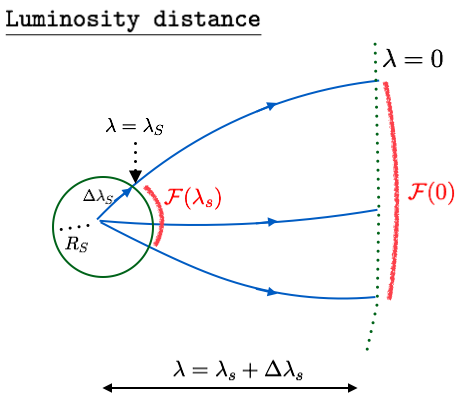}
  \includegraphics[width = 0.04 \textwidth]{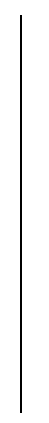}
  \includegraphics[width = 0.41 \textwidth]{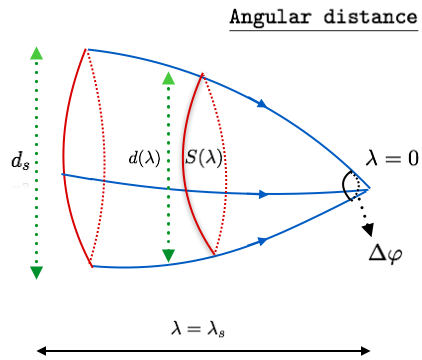}
 \caption{\it Representative plot of  the GW rays from source to detector position. The path of the GW bundle in blue is parameterized by the affine parameter $\lambda$. {\bf Left panel}: quantities
 entering the GW luminosity distance, associated with a GW bundle diverging from source to detector. {\bf Right panel}: quantities
 entering the GW angular distance, associated with a GW bundle converging from source to detector. See text for definitions.}
 \label{fig:lumidis}
\end{figure}

Our treatment  in this section follows very closely the discussion of the classic paper by Sasaki
\cite{Sasaki:1987ad} that for the first time discussed the concept of luminosity and
angular distances for photons propagating in a perturbed FRW universe. Sasaki's early work
was followed by many articles that further generalized it extending the analysis
of luminosity distance for photons in a perturbed background -- see e.g. \cite{Pyne:2003bn,Holz:2004xx,Bonvin:2005ps,Hui:2005nm,Laguna:2009re,Bertacca:2017vod,Fier:2021fbt}.
Nevertheless, as we are going to discuss,  the formalism developed in  \cite{Sasaki:1987ad}  is
sufficiently flexible  to be applied to GW propagation on our scalar-tensor systems,
 with little adaptations needed along the way.

 \bigskip

 We start by introducing some geometric tools  we need for our arguments.  The space-time metric we are interested in is written as a conformally flat FRW universe, and reads
\be
d s^2\,=\,a^2(\eta) \,g_{\mu\nu} d x^\mu d x^\nu\,=\,\hat g_{\mu\nu}\,d x^\mu d x^\nu\,,
\ee
where from now on in this section we denote with a hat $\hat g_{\mu\nu}$ the physical space-time metric, while $ g_{\mu\nu}$ is
 the `comoving' part of the metric tensor.
 Analogously to  eq \eqref{dec1g}, we can write
\be
g_{\mu\nu}\,=\,\bar g_{\mu\nu}+h_{\mu\nu}\,,
\ee
and $h_{\mu\nu}$ the high-frequency field. In this cosmological context, the comoving metric components
$\bar g_{\mu\nu}$ corresponds to  Minkowski space-time,  plus the long wavelength perturbations.
Gravitational waves correspond to transverse-traceless metric fluctuations, whose null-like
4-momentum propagates along null geodesics.
The physical GW energy momentum tensor is given by eq \eqref{tmngwb}:

\be\label{tmngwb2}
\hat T^{\mu\nu}\,=\,\frac{e^{-\int {\cal T}}}{32\,\pi}\,{\cal A}_T^2\,\hat k^\mu \hat k^\nu\,,
\ee
where the overall factor depending on ${\cal T}$ controls the modified
gravity contribution, and the GW physical 4-momentum is
\be
\hat k^\mu\,=\,\frac{d x^\mu}{d \hat \lambda}\,,
\ee
with $\hat \lambda$ the physical affine parameter along the GW ray, as associated with the vector $\hat k^\mu$.
 Following the procedure developed in section \ref{secGeoAN}, we conveniently express
the evolution eq for the GW amplitude ${\cal A}_T$ as
 \be\label{evAtph}
\frac{d\,( e^{-\frac12\int {\cal T}} {\cal A}_T)}{d \hat \lambda }+\frac12\,( e^{-\frac12\int {\cal T}} {\cal A}_T)\,\hat \theta\,=\,0 \hskip1cm{\rm with} \hskip1cm
\hat \nabla_\mu \hat k^\mu\,=\,\hat \theta\,,
\ee
where $\hat \theta$ is the expansion parameter along the GW geodesics.

It is convenient to pass from physical (hat)  to comoving (no hat) quantities.
A conformal transformation maps a null GW geodesics in $\hat g_{\mu\nu}$ into null geodesics in $ g_{\mu\nu}$
\cite{Wald:1984rg}. The GW
affine parameter scales as $$d \hat  \lambda \to d \lambda\,=\,a^{-2}\,d \hat \lambda\,,$$ with $a$ the scale factor. The evolution equation in the comoving frame then results
\be\label{evecof1}
\frac{d\,( e^{-\frac12\int {\cal T}} {\cal A}_T\,a)}{d \lambda }+\frac12\,( e^{-\frac12\int {\cal T}} {\cal A}_T\,a)\, \theta\,=\,0 \hskip1cm{\rm with} \hskip1cm
 \nabla_\mu  k^\mu\,=\, \theta\,.
\ee

The evolution equation for the comoving expansion parameter $\theta$ can be
easily determined \cite{Sasaki:1987ad}, finding
\bea
\label{evoex1}
\frac{d \,\theta}{d\,\lambda}
&=&-R_{\mu\nu} k^\mu k^\nu-\frac{\theta^2}{2}-2\sigma^2\,,
\eea
where $\sigma$ is the shear along the GW geodesics:
\be
\sigma^2\,=\,\frac12 k_{(\alpha;\,\beta)}\,k^{(\alpha;\,\beta)}-\frac{\theta^2}{4}\,,
\ee
and $R_{\mu\nu} $ the space-time Ricci tensor. The graviton number conservation  \eqref{evSAv2} remains unchanged and  reads
\be  \label{evSAL}
\frac{d}{d \lambda}
\left\{
 e^{-\int {\cal T}}\,
 {\cal A}_T^2\, S
 \right\}
\,=\,0
\ee
in this comoving frame ($\lambda$ referring to the comoving affine parameter). These are the geometric ingredients we need for the physical
considerations we develop next.

\subsection{The GW luminosity distance}
\label{sec-dL}

To determine the  GW luminosity distance we proceed step-by-step as  \cite{Sasaki:1987ad}.
We introduce an observer whose physical four-velocity we denote with $\hat u^\mu$.
 The measured GW energy flux  by such observer
reads
\bea
\hat{ \cal F}^\alpha&=&-\hat T^{\mu}_{\,\,\nu}\,h^{\alpha}_{\,\,\mu}\,\hat u^\nu\,,
\\
&=&{\cal F}\,\hat n^\alpha\,,
\eea
where $\hat T^{\mu}_{\,\,\nu}$ is the GW energy momentum tensor \eqref{tmngwb2},
and
 \bea
h^{\alpha}_{\,\,\mu}&=&\delta^{\alpha}_{\,\,\mu}+ \hat u^\alpha\,\hat u_\mu
\,,
\\
\hat n^\alpha&=&\frac{1}{\omega}\left( \hat k^\alpha-\omega \hat u^\alpha\right)
\,,
\eea
while the GW flux amplitude and GW frequency measured by the observer
are
\bea
\label{fluxd1}
{\cal F}&=&\frac{e^{-\int\,{\cal T}}}{32\,\pi}\,{\cal A}_T^2\,\omega^2\,,
\\
\omega&=&-\hat k_\mu  \hat u^\mu\,.
\eea
The notion of GW frequency allows us to define the GW redshift $z$ at  the value $\lambda$
 of the comoving  GW geodesics affine parameter $\lambda$

\be\label{defred}
1+z(\lambda)\,=\,\frac{\omega (\lambda)}{\omega(0)}\,.
\ee

For computing the luminosity distance, we assume that GW are emitted by an approximately spherically
symmetric system, with characteristic radius $R_s$ (this assumption is nevertheless not important since
we send $R_s\to0$ at the end of the calculation). The flux amplitude ${\cal F}$ measured at the source
position  is related with the intrinsic source luminosity by the relation
\be
{\cal F}(\lambda_s)\,=\,\frac{{\cal L}_{GW}}{4 \pi\,R_s^2}\,,
\ee
with $\lambda_s$ the conformal affine parameter at the source. See Fig \ref{fig:lumidis}, left panel.  The luminosity distance to the source
as measured by an observer located at $\lambda=0$ is {\it defined} as
\be
d_L^{\rm (GW)}\,\equiv\,\left[\frac{{\cal L}_{GW}}{4 \pi\,{\cal F}(0)}\right]^{\frac12}\,=\,\sqrt{\frac{{\cal F}(\lambda_s)}{{\cal F}(0)}}\,R_s\,.
\ee
Substituting relation \eqref{fluxd1}, we find the following expression
 \be \label{expLD1}
d_L^{\rm (GW)}\,=\,\left(
e^{-\frac12\,\int_0^{\lambda_s} \,{\cal T}}
 \right)\,
\frac{{\cal A}_T(\lambda_s)}{{\cal A}_T(0)}\,\left[1+z (\lambda_s)\right]\,R_s\,.
\ee
Notice the role of the modified gravity friction term in the overall exponential factor, containing the
cumulative integral of the friction parameter ${\cal T}$ along the GW geodesics path (see
the discussion after eq \eqref{defoI}). Before proceeding, to make
contact with the literature, it is interesting to consider the ratio between the GW luminosity distance
\eqref{expLD1} versus the electromagnetic luminosity distance. This ratio provides an interesting
observable in case of multimessenger events. We find
\be \label{ratiolu1}
\frac{d_L^{\rm (GW)}}{d_L^{\rm (EM)}}\,=\,
\exp{\left[-\frac12\,\int^{\lambda_S}_{0}\,{\cal T}\,\frac{d \bar \phi}{d \lambda'}\,d \lambda' \right]}
\,,
\ee
singling out the modified gravity contribution as an integral from $\lambda=0$ (the position of the observer) to the source at $\lambda=\lambda_S$.  Once substituting in the integrand of \eqref{ratiolu1} the explicit
form of friction terms used for parameterizing deviations from GR in cosmological models
of dark energy, one finds exactly the same formulas used in the literature -- see
  Appendix
\ref{app-compLISA} for a discussion of such comparison.

\bigskip

We now proceed expressing the GW luminosity distance \eqref{expLD1} in an alternative way, that is more
useful for explicitly including  effects of cosmological perturbations, and for then
 comparing with the angular distance in section \ref{sec-dA}. From now on, we denote a
 a perturbed quantity with a tilde, and unperturbed without tilde. For example, we write for the comoving metric
\be
\tilde g_{\mu\nu}\,=\,\eta_{\mu\nu}+\delta g_{\mu\nu}\,,
\ee
meaning that $\delta g_{\mu\nu}$ are long-wavelength perturbations.
We now introduce a null vector $\tilde K^\mu$   proportional to $\tilde k^\mu$, and  use it
 to define a corresponding   affine parameterization:
\be
\tilde K^\mu\,\equiv\,-\frac{\tilde k^\mu}{\tilde \omega(\lambda_s) a(\tilde \eta (\lambda_s))}
\,,
\ee
where $\lambda$ from now on is the affine parameter associated with $\tilde K^\mu$.
This vector  is normalized
in such a way
that, once evaluated  at the source $\lambda=\lambda_s$, we find
\be
\left( \tilde g_{\mu\nu} \tilde K^\mu \tilde u^\nu \right)_{\lambda_s}\,=\,1\,,
\ee
where $ \tilde u^\mu $ is the perturbed comoving observer 4-velocity.

 The introduction of the vector $\tilde K^\mu$. is technically convenient to easily relate the physical size of the source with the
 affine parameter along the GW geodesics. In fact,
  as shown in \cite{Sasaki:1987ad}, the characteristic
size $R_s$ of the source can be expressed as
\be
R_s\,=\,a(\tilde \eta (\lambda_s))\,\Delta \lambda_s\,,
\ee
with $\Delta \lambda_s$
the infinitesimal affine parameter associated with the source size. See Fig \ref{fig:lumidis}, left panel.

 If the unperturbed case,
 it is straightforward to integrate \eqref{evoex1} (recalling that we have no shear for a spherically
symmetric source): we get
\be
\theta_L\,=\,\frac{2}{\lambda-\lambda_s-\Delta \lambda_s}\,,
\ee
where the suffix $L$ is included to associate the expansion parameter $\theta$ with the
 luminosity distance.
  The deviation for the expression of $\theta_L$ at first order  in cosmological  inhomogeneities can be expressed
as
\be
\delta \theta_L(\lambda)\,=\,\tilde  \theta_L\left[ x^\mu(\lambda)+\delta x^\mu (\lambda)\right]-
  \theta_L\left[ x^\mu(\lambda)\right]\,.
\ee
The evolution equation for the first order perturbation $\delta \theta_L(\lambda)$ can be obtained from
from  \eqref{evoex1}. It  reads
\be\label{evexp2p}
\frac{d\, \delta \theta_L}{d \lambda}\,=\,-\theta_L\,\delta \theta_L-\delta (R_{\mu\nu} K^\mu K^\nu)_\lambda
\ee
where $R_{\mu\nu}$ is the perturbed space-time Ricci tensor at the position $\lambda$ along the GW geodesics.
Integrating eq \eqref{evexp2p} along the  affine parameter,
 imposing the boundary condition $ \delta \theta_L(\lambda_s)=0$,  we get
\be
\delta  \theta_L(\lambda)\,=\,\frac{1}{(\lambda-\lambda_s -\Delta \lambda_s)^2}\,\int_\lambda^{\lambda_s}
\, d \lambda'\,\left( \lambda'-\lambda_s  -\Delta \lambda_s\right)^2 \,\delta (R_{\mu\nu} K^\mu K^\nu)_{\lambda'}
\,\,.
\ee
Collecting all the results so far, we can integrate eq \eqref{evecof1}, and get the relation
\be\label{defLS1a}
\exp{\left[-\frac12\,\int_0^{\lambda_s}\,{\cal T} \right]}\,\times\,\frac{{\cal A}_T (\lambda_s)\,a(\tilde \eta (\lambda_s))}{{\cal A}_T (0)\,a(\tilde \eta (0))}\,=\,\frac{\lambda_s+\Delta \lambda_s}{ \Delta \lambda_s}\,\exp{\left[ -\frac12 \int_0^{\lambda_s}
d \lambda\,\delta\theta_L (\lambda)
\right]}\,\,.
\ee
This result can be inserted into
eq \eqref{expLD1}: taking $\Delta \lambda_s\to0$ (i.e. considering a source of negligible size)  we end with the compact expression

\be \label{finpDLGW}
\tilde d_L^{\rm(GW)}(\lambda_s)\,=\,\lambda_s\,a[\tilde \eta(0)]\,\left[1+\tilde z(\lambda_s) \right]\,\times\,\exp{\left[ -\frac12 \int_0^{\lambda_s}
d \lambda\,\delta\theta_L (\lambda)
\right]}\,.
\ee

\bigskip

\noindent
Notice that all the effects of modified gravity friction term are implicitly included in the expression \eqref{defLS1a}, which  relates the affine parameter $\lambda_s$ with the remaining quantities.

The compact expression  \eqref{finpDLGW} (accompanied by relation \eqref{defLS1a})
is exact and include the effects of cosmological fluctuations -- on the other hand is implicitly expressed
in terms of  $\lambda_s$, and is not easy from it to extract in a physically transparent way the implications of cosmological fluctuations and of modified gravity. Such implications are more easily studied by using  the cosmic
ruler formalism -- see \cite{Garoffolo:2019mna}. This approach explicitly identifies contributions from  peculiar velocities,
weak lensing, Sachs-Wolfe effects, volume effects, and Shapiro time delay, and allows to appreciate
the contributions due to modified gravity. We refer the reader to   \cite{Garoffolo:2019mna} for more details:
for our purposes to prove
 the validity of Etherington reciprocity law    -- our aim for the next section -- formula \eqref{finpDLGW} will be sufficient.

\subsection{The GW angular distance, and Etherington reciprocity law}
\label{sec-dA}

We now prove the validity of Etherington reciprocity law connecting luminosity and angular GW distances. This
relation is expected to hold in scenarios where graviton number is conserved, as our scalar-tensor set-up (see
section \ref{sec-consgn}).

The GW angular distance $d_{A}^{\rm GW}$ is formally defined in terms of the ratio between  the angular diameter $d_{s}$ of the source located at conformal affine parameter $\lambda_s$,
and  the source apparent angular size $\Delta \phi$ as measured by an observer at $\lambda=0$.
In formulas:
\be
d_{A}^{\rm (GW)}\,=\,\frac{d_{s}}{\Delta \phi}\,.
\ee
 Following \cite{Sasaki:1987ad},
it is convenient to re-express $d_{A}^{\rm ( GW)}$ as
\be \label{defDA2}
d_A^{\rm (GW)}\,=\,\left( \frac{S(\lambda_s)}{S(\Delta \lambda)} \right)^{1/2}\,\frac{{\bf d}(\Delta \lambda)}{\Delta \phi}
\,,
\ee
with $S(\lambda)$ the cross-section area of GW rays at $\lambda$, and its diameter by ${\bf d}(\lambda)$.
 $\Delta \lambda$ is  the affine parameter in proximity of the observer.
  \cite{Sasaki:1987ad} proved the relation
\be \label{sasaA2}
\frac{{\bf d}(\Delta \lambda)}{\Delta \phi}\,=\,\frac{a^2[\tilde \eta(0)]\,\Delta \lambda}{(1+\tilde z(\lambda_s)) \,a[\tilde \eta(\lambda_s)]}
\,,
\ee
connecting the ratio ${{\bf d}(\Delta \lambda)}/{\Delta \phi}$ with  $\Delta \lambda$.  See Fig \ref{fig:lumidis}, right panel.

  Fig \ref{fig:lumidis} shows that in evaluating the angular
 distance one considers GW bundles expanding from the observer position (while, on the contrary,  the luminosity
 distance considers bundles expanding from the source). Hence, the expansion parameter $\theta_A$
 associated with angular distance reads, when neglecting effects of cosmological inhomogenities, can be obtained
  integrating eq \eqref{evoex1}:
 \be \label{intthA}
\theta_A\,=\,\frac{2}{\lambda}
\,.
\ee
When including the contributions of
 perturbations, we find the formal solution
\be
\delta \theta_A\,=\,-\frac{1}{\lambda^2}\,\int_0^\lambda\,d \lambda'\,\lambda'^2\,\delta(R_{\mu\nu} K^\mu K^\nu )_\lambda'\,,
\ee
for the first order perturbation $\delta \theta_A$ to the angular expansion parameter.

Integrating  eq \eqref{evSAL} along the GW geodesics, and comparing with the definition \eqref{defDA2}, we get the
relation

\be\label{defAD2}
d_A^{\rm (GW)}\,=\, \exp{\left[\frac12\,\int_0^{\lambda_s}\,{\cal T} \right]}\,\times\,
 \left(
 \frac{{\cal A}_T(0)}{
 {\cal A}_T(\lambda_s)}
 \right)\,\times\,
\frac{{\bf d}(\Delta \lambda)}{\Delta \phi}
\,.
\ee

Moreover, integrating 
 eq \eqref{evecof1} using the result \eqref{intthA}, we now obtain
\be\label{defLS1}
\exp{\left[-\frac12\,\int_0^{\lambda_s}\,{\cal T} \right]}\,\times\,\frac{{\cal A}_T (\lambda_s)\,a(\tilde \eta (\lambda_s))}{{\cal A}_T (0)\,a(\tilde \eta (0))}\,=\,\frac{\Delta \lambda}{  \lambda_s}\,\exp{\left[ -\frac12 \int_{\Delta \lambda}^{\lambda_s}
d \lambda\,\delta\theta_A (\lambda)
\right]}
\ee
as the relation between affine parameter and angular expansion parameter.
 Substituting the results of eqs \eqref{defLS1} and \eqref{sasaA2} into eq \eqref{defAD2}, we obtain
  the expression
 \be
 \tilde d_A^{\rm (GW)}\,=\,\frac{a[\tilde \eta(0)]}{1+\tilde z(\lambda_s)}\,\lambda_s\,
 \exp{
 \left[
 \frac12 \int_0^{\lambda_s}\,
 \delta \theta_A(\lambda) d\lambda
 \right]}\,.
 \ee
 Comparing with the expression for the luminosity distance, we get
 \bea
\label{forAp1}
 \tilde d_A^{(\rm GW)}&=&\frac{\tilde d_L^{\rm (GW)}}{(1+\tilde z)^2}\, \exp{
 \left[
 \frac12 \int_0^{\lambda_s}\,
\left( \delta \theta_A(\lambda)+\delta \theta_L(\lambda) \right)d\lambda
 \right]}\,,
 \\
 \label{ether1}
 &=&\frac{\tilde d_L^{\rm (GW)}}{(1+\tilde z)^2}\,.
 \eea
The second line, eq \eqref{ether1}, is the desired Etherington relation, valid including first order perturbations.
(The step between eq \eqref{forAp1} and \eqref{ether1} requires  technical calculations that we defer to
 Appendix \ref{appEthP}.)

 \smallskip

 Hence we proved that in the scalar-tensor framework discussed in this work, with conservation of graviton
 number, luminosity and angular distances for GW are connected by the classic Etherington law \eqref{ether1}.
  We have seen that GW and electromagnetic luminosity distances can differ -- see eq \eqref{ratiolu1} -- and
 this fact is important in case of multimessenger events. Then eq  \eqref{ether1} tells us that  the same is true for
 angular distances, and we can schematically write a relation analog to eq \eqref{ratiolu1}:

\be \label{ratioan1}
\frac{d_A^{\rm (GW)}}{d_A^{\rm (EM)}}\,=\,\exp{\left[-\frac12\,\int^{\lambda_S}_{0}\,{\cal T}\,\frac{d \bar \phi}{d \lambda}\,d \lambda \right]}\,.
\ee
In what comes next we briefly discuss some applications of these results to   GW lensing.

\subsection{Implications for GW lensing}
\label{subsec-GWlensing}

Strong
GW lensing from large-scale structures  between GW source and detector is an important
phenomenon that -- although not yet observed --  is likely to offer new ways to probe
cosmological parameters with future gravitational wave detections. For example LISA, by
 observing sources from high-redshift sources,  will likely detect lensed events \cite{Oguri:2018muv}.
See e.g.
 \cite{Turner:1990mk,Wang:1996as,Nakamura:1997sw,Wang:1999me,Macquart:2004sh,Takahashi:2003ix,Seto:2003iw,Yoo:2006gn,Seto:2009bf,Shapiro:2009sr,Sereno:2010dr,VanDenBroeck:2010fp,
Sereno:2011ty,Baker:2016reh,Kyutoku:2016zxn,Collett:2016dey,Dai:2017huk,Oguri:2018muv,Dai:2018enj,Haris:2018vmn,Oguri:2018muv,Hou:2019dcm,Liu:2019dds,Meena:2019ate,Hannuksela:2019kle,Morita:2019sau,
Cusin:2020ezb,Pang:2020qow,Rubin:2020gba,Ezquiaga:2020dao,Ezquiaga:2020gdt,Ezquiaga:2020spg,Takahashi:2016jom,Suyama:2020lbf,Contigiani:2020yyc}
   for works discussing
this topic.

We consider strong GW lensing from point-like lenses in the geometric optics limit, valid
when the GW wavelength is well shorter than the Schwarzschild radius of the lens. In this limit, we do not
need to discuss interference effects that, although very interesting, go  beyond the scope of this
work.   We focus on the specific observable associated with the
 {\it time-delay} that the presence of the lens induces on the
propagation time of the GW from source to detector.  We compare the GW
time-delay induced by the presence of the lens with the electromagnetic (EM) time
delay of lensed light received in a multimessenger detection.

The works \cite{Suyama:2020lbf,Ezquiaga:2020spg} shown conclusively that GW and EM lensed signals arrive
at the same time at the detector, if both waves propagate at the same speed and are emitted
at the same time. In the geometric optics limit this is expected when photons and GW travel through null geodesics, since
by definition both sectors cover the minimal possible distance from source to detector.
 Causality arguments based on Fermat principle allow one to prove this statement in full
generality. \cite{Suyama:2020lbf} also argues that the same result should be valid in any theory of
gravity, to respect causality.

Said this, it is interesting to analyze the topic in an explicit
modified gravity set-up, for understanding how effects of modified gravity balance so to ensure
the same time delay for GW and light. This is the scope of this section. We find this topic interesting
since the 
expression for the time-delay commonly used in the literature  (see e.g. \cite{schneider}, as well
as the recent \cite{Ezquiaga:2020gdt} in the context of gravitational waves) explicitly contains
factors depending on the angular distance $d_A^{\rm {(GW)}}$, which can be modified
with respect to the standard case (see eq \eqref{ratioan1}).
 In fact, the GW time delay $\Delta t^{\rm (GW)}$
 has a geometrical contribution, and a Shapiro contribution $t_\Phi^{\rm (GW)}$ due to the presence of inhomogeneities in the background cosmological
space-time crossed by  GW in their path from source to detection. We express it in the following form
  \be \label{genexTD}
\Delta t^{\rm (GW)}\,=\,(1+z)\,\frac{d_{OL}^{\rm (GW)} d^{\rm (GW)}_{SO}}{2\, d^{\rm (GW)}_{SL}}\,|\theta-\theta_S\,|^2+ t_\Phi^{\rm (GW)}\,.
\ee
In the previous expression, $z$ is the redshift, $d_{OL}^{\rm (GW)}$  the GW angular distance as measured
from the observer to the lens,  $d^{\rm (GW)}_{SO}$ the same quantity measured
from source to the observer, and $d^{\rm (GW)}_{SL}$ from source to lens. $t_\Phi^{\rm (GW)}$  is
the aforementioned  Shapiro contribution due to inhomogeneities.
$\theta$ is the observed angular position of the source,  $\theta_S$ the would-be angular position of the source in absence of the lens.
In the electromagnetic case, the
corresponding time-delay $\Delta t^{\rm (EM)}$ has exactly the same structure, changing the suffixes
from GW to EM. See Appendix \ref{app-timedelay} for a derivation of the geometric part (the first term) of the previous formula.

In \cite{Garoffolo:2019mna} we shown explicitly that the Shapiro time-delay $t_\Phi$ is exactly the same for GW and
EM observations in a scalar-tensor framework, $t_\Phi^{\rm (GW)}=t_\Phi^{\rm (EM)}$: we refer the reader to this work for full details. However,
 eq \eqref{genexTD} also contains explicitly angular distances in the first geometric term, which
 we dub $\Delta t_{\rm geo}$.
  Using eq  \eqref{ratioan1} we can in fact understand how
 the effects of modified gravity in the geometrical time-delay  $\Delta t_{\rm geo}$ compensate.
 We can write:
\bea
\Delta t_{\rm geo}^{\rm (GW)}&=&
(1+z)\,\frac{d_{OL}^{\rm (GW)} d^{\rm (GW)}_{SO}}{2\, d^{\rm (GW)}_{SL}}\,|\theta-\theta_S\,|^2
\,,
\nonumber
\\
&=&\left( \frac{d_{OL}^{\rm (GW)}}{d_{OL}^{\rm (EM)}} \right)
\left( \frac{d_{SO}^{\rm (GW)}}{d_{SO}^{\rm (EM)}} \right)
\left( \frac{d_{SL}^{\rm (EM)}}{d_{SL}^{\rm (GW)}} \right)\,\Delta t_{\rm geo}^{\rm (EM)}
\,,
\nonumber
\\
&=&\left( \exp{\left[ -\frac12
\int_{\lambda_O}^{\lambda_L} {\cal T} \,\frac{d \bar \phi}{ d \lambda}\,d \lambda
 -\frac12
\int_{\lambda_S}^{\lambda_O} {\cal T} \,\frac{d \bar \phi}{ d \lambda}\,d \lambda
+\frac12
\int_{\lambda_S}^{\lambda_L} {\cal T} \,\frac{d \bar \phi}{ d \lambda}\,d \lambda
\right]}
\right)  \,\Delta t_{\rm geo}^{\rm (EM)}
\,,
\nonumber
\\
&=&\,\Delta t_{\rm geo}^{\rm (EM)}\,.
\label{eqtims1}
 \eea
To pass from the second to third line we used relation \eqref{ratioan1}, while the fourth line is
a simple consequence of  summing the integrals in the exponent. So we find that modified gravity contributions
carefully compensate, leading to the equality $\Delta t_{\rm geo}^{\rm (GW)}\,=\,\Delta t_{\rm geo}^{\rm (EM)}$. Together with the fact that the Shapiro time delay coincides in the GW and EM sector, we find
 that by using the expression eq \eqref{genexTD}
we are ensured that
 GW and EM experience the same time-delay from strong lensing in a scalar-tensor theory of gravity,
as we wish to demonstrate.  In particular, even if formula \eqref{genexTD} is expressed in terms of angular distances that, when
taken individually,  can be  different with respect to General Relativity,  one finds cancelations between
modified gravity coefficients that lead to  the  result \eqref{eqtims1}.

\section{Conclusions}
\label{sec-conc}

In this work we studied
the propagation of high-frequency gravitational waves (GW) in scalar-tensor theories of gravity, with the aim of examining   properties of cosmological distances as inferred from GW measurements.
We first developed a covariant set-up for  our scalar-tensor systems,  which by hypothesis are
         characterized by
      symmetry properties based
     on coordinate invariance. Symmetry considerations allowed us to
 extract  transverse-traceless components of the high-frequency scalar-tensor fluctuations, identified with GW. In scenarios where scalar and tensor components propagate with different speeds the two
 sectors decouple at the linearized level around an arbitrary background, and the evolution of high-frequency GW and scalar modes  can be  studied independently.

 We then  determined the most general structure of the   GW linearized equations and of the GW energy momentum tensor,  assuming  that GW move with the speed of light.    Modified gravity effects are encoded in a small number of parameters, and we studied the conditions for ensuring graviton number conservation in our set-up.
We  then applied our general findings to the case of   GW propagating   through a perturbed  cosmological space-time, deriving the expressions  for the GW luminosity distance  $d_L^{({\rm GW})}$ and  the GW angular distance $d_A^{({\rm GW})}$.
Both luminosity and angular
 distances can be modified with respect to General Relativity.
We    proved for the first time the validity of Etherington reciprocity law  $d_L^{({\rm GW})}\,=\,(1+z)^2\,d_A^{({\rm GW})}$ in a perturbed universe   within a scalar-tensor framework.   We  discussed  implications  of this result for  gravitational lensing, focussing   on  time-delays of lensed GW and lensed photons emitted simultaneously   in   a multimessenger event.   We explicitly
wrote an expression for the time-delay formula,
showing how modified gravity effects carefully compensate between different   contributions.
   As a result,   lensed GW   arrive at  the same  time as     their lensed electromagnetic  counterpart, in agreement with causality constraints.

It would be interesting to consider scenarios where graviton number is not conserved, as done in alternative theories of cosmology where the photon number is not conserved (see e.g.
 \cite{Bassett:2003vu}). 
 Another avenue for future research is to extend our arguments to scenarios
 with non-standard dispersion relations where 
   the speed of tensor modes is not equal to  one, at least in some frequency ranges.
 More in general, it would be interesting to explore in more details
 how observables
 depending on the GW angular distance and GW lensing can be used to probe alternative theories of gravity. 
 We plan to study these topics in  future works.


    \subsection*{Acknowledgments}

It is a pleasure to thank Carmelita Carbone for input and discussions. 
GT is partially funded by  the STFC grant ST/T000813/1.
AG acknowledges support from the NWO and the Dutch Ministry of Education,
Culture and Science (OCW) (through NWO VIDI Grant No. 2019/ENW/00678104 and from the D-ITP consortium).
 DB and SM acknowledge partial financial support by ASI Grant No. 2016-24-H.0 and funding from Italian Ministry of Education, University and Research (MIUR) through the Dipartimenti di eccellenza project ``Science of the Universe".


\begin{appendix}
\section{A covariant approach to high-frequency fluctuations}\label{sec_theory1}

In the main text we derived the covariant evolution equations and the
energy-momentum-tensor for transverse-traceless GW excitations. In this appendix
we spell out the technical arguments   used
for identifying the GW sector, and for distinguishing it from the scalar sector. We work within
 the covariant approach of  section \ref{sec-ourset}.

\subsection{Decomposing the gauge transformations}

Assuming a time-like~\footnote{The case of space-like direction can
be studied with little changes by the same approach. We assume nevertheless that $v^\mu$ never
becomes null-like.} direction $v_\mu$ for  the scalar gradient of eq \eqref{defgrad1}, we introduce the vector
\be\label{defXmu1}
X_\mu \equiv \frac{v_\mu}{ \sqrt{ 2 X}} \hskip1cm \mbox{such that}\hskip1cm X^\mu X_\mu = -1\,
\ee
where
\be
X \equiv - (v^\mu v_\mu) / 2
\,.\ee
We decompose the gauge vector $\xi_\mu$ in eqs~\eqref{trasf1}, \eqref{trasf2}
in a part orthogonal, and a part parallel to $v^\mu$ :
\be\label{gausplit}
\xi_\mu\,=\,\xi_\mu^{(T)}+ X_\mu\,\xi^{(S)} \hskip1cm,\hskip1cm  X^\mu\,\xi_\mu^{(T)}\,=\,0\,.
\ee
This decomposition defines what we call a {\it $T$-gauge transformation},  proportional to the vector $\xi_\mu^{(T)}$
orthogonal to $v^\mu$, and
an a {\it  $S$-gauge transformation}, depending on the scalar $\xi^{(S)}$.
The metric and scalar field perturbations transform under a gauge transformation as
\bea
h'_{\mu\nu} &=& h_{\mu\nu} - \big( \nabla_\mu \xi_\nu +  \nabla_\nu \xi_\mu\big) \,, \\
\varphi' &=& \varphi - v^\mu \,\big(\xi^{(T)}_\mu + X_\mu \xi^{(S)}\big) = \varphi + \sqrt{2 X} \,\xi^{(S)} \,.
\eea
Hence, $\varphi$ transform only under an $S$-gauge transformations, while $h_{\mu\nu}$ both $S$ and $T$-gauge transformations. We now show how to  use the decomposition \eqref{gausplit} to consistently distinguish GW from scalar excitations in model-independent framework.

We introduce the quantity
\be\label{TildeH}
\tilde h_{\mu\nu} \equiv h_{\mu\nu} +  \nabla_\mu H_\nu +   \nabla_\nu H_\mu\,. \hskip1cm {\text{with}}\hskip1cm  H_\mu \equiv \frac{X_\mu}{\sqrt{2 X}} \, \varphi\,.
\ee
Each of the contributions to $\tilde h_{\mu\nu} $  are of the same order in the gradient expansion $\epsilon$,
since we are assuming ${\cal O}(\varphi) \sim \epsilon \,{\cal O}(h)$.
The field combination in eq.\eqref{TildeH} is  $S$-gauge invariant:
\bea \label{TildeHtransf}
\tilde h'_{\mu\nu} = \tilde h_{\mu\nu} -  \nabla_\mu \, \xi^{(T)}_\nu -  \nabla_\nu  \, \xi^{(T)}_\mu \,.
\eea
We define the orthogonal projection operator relative to the vector $X^\mu$
\be\label{projector}
\Lambda_{\mu \nu } = \bar g_{\mu\nu} + X_\mu X_\nu \,,
\ee
such  that $ X^\mu  \Lambda_{\mu\nu} = 0$, and we apply it to $\tilde h_{\mu\nu}$. We find
\bea
\tilde h_{\mu\nu} = X_\mu X_\nu \,  h^{(S)} - \left( X_\mu  h^{(V)}_\nu + X_\nu  h^{(V)}_\mu \right) +  h^{(T)}_{\mu\nu}\,.
\eea
The quantities
\be\label{defcom123}
 h^{(S)} \equiv  X^\rho \, X^\sigma \,\tilde h_{\rho \sigma} \,, \qquad\qquad
 h^{(V)}_\mu \equiv X^\rho\, \Lambda^\sigma_\mu \,\tilde h_{\rho \sigma} \,, \qquad\qquad
 h^{(T)}_{\mu\nu} \equiv \Lambda^\rho_\mu\,  \Lambda^\sigma_\nu \,\tilde h_{\rho \sigma} \,,
\ee
 are  $S$-gauge invariant since $\tilde h_{\mu\nu}$ is so. On the
other hand, they can transform under $T$-type transformations. Under the gauge transformation~\eqref{TildeHtransf}, the quantities $ h_{\dots}^{(S,\,V,\,T)}$ can develop contributions at order ${\cal O}(\epsilon^{0})$ and ${\cal O}(\epsilon^{1})$ because ${\cal O}(\xi^{(T)}_\mu) \, \sim \,\epsilon \, {\cal O}(h_{\mu\nu})$. Since we are interested in characterizing metric fluctuations up to order ${\cal O}(\epsilon^{0})$ -- see discussion around eq.~\eqref{ampscal1} -- we  neglect  contributions at order ${\cal O}(\epsilon^{1})$. Moreover, by construction, we have the orthogonality property
\be
X^\mu \,  h^{(V)}_\mu = X^\mu  h^{(T)}_{\mu\nu} = 0\,.
\ee

Therefore, using eq.~\eqref{TildeHtransf}, we find that, under a $T$-gauge transformation and up to ${\cal O}(\epsilon^{0})$, the quantities  $ h_{\dots}^{(S,\,V,\,T)}$ transform as
\bea
 h^{'(S)} &=&  h^{(S)} \,,\\
 h^{'(V)}_{\mu} &=&   h^{(V)}_{\mu} - X^\rho \,  \nabla_\rho \, \xi^{(T)}_\mu \,, \label{TildeHVtransf}\\
 h^{'(T)}_{\mu\nu} &=&  h^{(T)}_{\mu\nu} - \left(  \nabla_\mu \xi^{(T)}_\nu +  \nabla_\nu \xi^{(T)}_\mu \right) - X^\rho \left( \, X_\mu \,  \nabla_\rho \, \xi^{(T)}_\nu  +  X_\nu \, \nabla_\rho \, \xi^{(T)}_\mu\,\right)\,, \label{TildeHTtransf}
\eea
thus, $   h^{(S)}$ is  also $T$-gauge invariant at order ${\cal O}(\epsilon^{0})$.

\subsection{Gauge fixing}
Since we demand that the linearized equations are  invariant under coordinate transformations -- i.e. separately $S$ and $T$-gauge invariant -- we can assume they can be organized in terms of the $S$-gauge invariant combinations $ h_{\dots}^{(S,\,V,\,T)}$. After identifying the gauge-invariant quantities, we can now make use of the $T$-gauge freedom of
eqs \eqref{TildeHVtransf} and \eqref{TildeHTtransf} for fixing convenient gauge conditions to study the physics of the system.

 The first gauge fixing condition we impose is
\be
 h^{'(V)}_{\mu} = 0\,, \label{Vgauge}
\ee
by choosing $\xi^{(T)}_\mu$ such that $ h^{(V)}_{\mu} = X^\rho  \nabla_\rho \, \xi^{(T)}_\mu$ in eq \eqref{TildeHVtransf}.
This condition is compatible, at order ${\cal O}(\epsilon^{0})$, with the orthogonality requirement   $X^\mu  h^{(V)}_{\mu}= 0$ since we have
\be
X^\mu X^\rho  \nabla_\rho \, \xi^{(T)}_\mu  = X^\rho  \nabla_\rho \, \left( X^\mu \, \xi^{(T)}_\mu \right) + {\cal O}(\epsilon^{1}) =  {\cal O}(\epsilon^{1})\,.
\ee
Eq.~\eqref{Vgauge} leaves the residual $T$-gauge freedom  $x_\mu \rightarrow x_\mu + \xi^{(T)}_\mu$, such that
\be \label{Vresidual}
X^\rho  \nabla_\rho \, \xi^{(T)}_\mu = 0\,.
\ee
We separate $ h^{(T)}_{\mu\nu}$ into a traceless plus trace components. Since $X^\mu  h^{(T)}_{\mu\nu} = 0$ we select the trace in the subspace orthogonal to $X^\mu$,
\be
 h^{(T)}_{\mu\nu} = \gamma_{\mu\nu} + \frac13 \, \Lambda_{\mu\nu} \, h^{\rm(tr)}
\ee
with
\bea
 h^{\rm(tr)} \equiv \bar g^{\mu\nu} \,  h^{(T)}_{\mu\nu} &=& \Lambda^{\mu\nu} \,  h^{(T)}_{\mu\nu}\,.
\eea
The quantity $\gamma_{\mu\nu}$ satisfies
\be
\Lambda^{\mu\nu} \gamma_{\mu\nu} = \bar g^{\mu\nu} \gamma_{\mu\nu} = 0
\ee
since $X^\mu \gamma_{\mu\nu} = X^\mu \left(  h^{(T)}_{\mu\nu} - \frac13 \, \Lambda_{\mu\nu} \, h^{\rm (tr)} \right) = 0$.
Starting from eq.~\eqref{TildeHTtransf}, we can find how $\gamma_{\mu\nu}$ and $ h^{\rm (tr)}$ transform under a gauge transformation. For those vector fields $\xi^{(T)}_\mu$ that satisfy eq \eqref{Vresidual} we have
\bea
  h^{'{\rm (tr)}} &=& h^{{\rm (tr)}} - 2  \nabla^\mu \, \xi^{(T)}_\mu \,,\\
\gamma'_{\mu\nu} &=& \gamma_{\mu\nu} -   \nabla_\mu \xi^{(T)}_\nu -  \nabla_\nu \xi^{(T)}_\mu  + \frac23 \, \Lambda_{\mu\nu} \,  \nabla^\rho\, \xi^{(T)}_\rho\,. \label{GammaTransf}
\eea
We now use eq.~\eqref{GammaTransf} to impose the transversality condition
 $\bar \nabla^\mu \gamma'_{\mu\nu} = 0$:
\be
\bar \nabla^\mu \gamma'_{\mu\nu}  = \bar \nabla^\mu \gamma_{\mu\nu}  - \bar \Box \, \xi^{(T)}_\nu  - \frac13 \nabla_\nu \left( \bar \nabla^\mu \xi^{(T)}_\mu \right)\,.
\ee
This  gauge transformation is valid up to order ${\cal O}(\epsilon^{-1})$.
We retain only contributions up to this order in $\epsilon$: this is  consistent with keeping only terms up  to ${\cal O}(\epsilon^{0})$ in   the gauge transformation of $\gamma_{\mu\nu}$, since  we are gauge fixing its gradient.
After such gauge choices, the quantity $\gamma'_{\mu\nu}$ is transverse and traceless. We dub it
\be
\gamma'_{\mu\nu}\,\equiv\,h_{\mu\nu}^{(TT)}\,,
\ee
 and  we identify it as the high-frequency GW discussed in the main text.
At this stage,
we point out that is not possible to choose $ h^{\rm (tr)} = 0$, within the residual gauge freedom given by~\eqref{Vresidual}, if $ h^{\rm (tr)}$ depends on the coordinate in the direction of  $X_\mu$. For simplicity,
we can exhaust the gauge freedom imposing $  \nabla^\mu \xi_\mu^{(T)} = 0$, such that the
trace   $ h^{\rm (tr)}$  is gauge-invariant, while the transverse-traceless
GW excitations $h_{\mu\nu}^{(TT)}$ are invariant under the  residual transformation that can be read from eq \eqref{GammaTransf}:
\be \label{resTTtr1}
h_{\mu\nu}^{(TT)}\to h_{\mu\nu}^{(TT)}-  \bar \nabla_\mu \xi^{(T)}_\nu - \bar \nabla_\nu \xi^{(T)}_\mu\,.
\ee

\medskip
To sum up, after imposing the  gauge conditions above,
  the $S$-gauge invariant metric perturbations \eqref{TildeH} we started with read
\be\label{TildeHfinal}
\tilde h_{\mu\nu} = X_\mu \, X_\nu \,  h^{(S)} + \frac13 \, \Lambda_{\mu\nu}  \,  h^{\rm (tr)} +  h_{\mu\nu}^{(TT)}\,.
\ee
We can also use the $S$-gauge for setting the unitary gauge $\varphi=0$: then,  after these gauge fixings,
$\tilde h_{\mu\nu} $ coincides with the original metric fluctuations $h_{\mu\nu}$.

\smallskip
Some words on the number of propagating degrees of freedom (dof).
The quantity $\tilde h_{\mu\nu}$, before we make any gauge choice, has $10$ non-vanishing components, each of them a potential dof. Making    gauge fixings as explained above we imposed $6$ conditions, since both  $ h^{(V)}_\mu$ and $ h_{\mu\nu}^{(TT)}$ are  by construction orthogonal to the vector $X^\mu$. Hence, we are left  with $4$ potential dof. In  section  \ref{appSDF} we show that only $3$ out of these $4$ are independent propagating dof, while $ h^{(S)}$ is a constrained field. The evolution
equations of the 3 propagating dof will be decoupled under physically reasonable assumptions
on the the velocities  of the fields involved.

\subsection{Separating the evolution equations}\label{appSDF}

We now discuss the evolution equations for the metric perturbation  of eq.~\eqref{TildeHfinal}, as
obtained from the linearized Einstein equations. We build arguments to show that at the linearized level, under physically reasonable conditions, different
sectors evolve independently one from the other.

Since by hypothesis our system is invariant under gauge diffeomorphisms, i.e. under coordinate
transformations, Einstein equations can be expressed in terms of the  fields
discussed above, obtained after gauge-fixing appropriate gauge-invariant quantities.
Therefore, in an unitary gauge with $\varphi=0$,
 we write the linearized Einstein equations as
\be\label{TildeEFE}
G^{(1)}_{\mu\nu} \left[ \tilde h_{\rho\sigma}\right] = T^{(1)}_{\mu\nu}\left[ \tilde h_{\rho\sigma} \right]\,,
\ee
with $\tilde h_{\rho\sigma}$ the $S$-gauge invariant combination given in eq \eqref{TildeHfinal}. Taking the trace
of the previous linear equation we eliminate from the left-hand-side the dependence on the transverse-traceless
fluctuation $ h^{(TT)}_{\mu\nu}$. In fact, the linearized Ricci scalar reads
\be \label{ricS1}
R^{(1)}\,=\,-\Box  h^{{\rm (tr)}} +  \Lambda^{\alpha \beta}\,\nabla_\alpha\nabla_\beta \left( h^{(S)} +  \frac13   h^{{\rm (tr)}}  \right)\,,
\ee
where $\Lambda_{\mu\nu}$ is the projector introduced in eq \eqref{projector}. We notice that while the trace scalar $h^{{\rm (tr)}}  $ receives a  kinetic contribution  controlled by the d'Alembertian operator $\Box$, second derivatives acting on the scalar  $  h^{(S)} $ are always weighted by the projector operator
 $\Lambda_{\mu\nu}$, hence they are directed on the space
 orthogonal to the vector $v^\mu$. Let us now
 use the  time-like vector $v^\mu$  for slicing the space-time into a family of space-like surfaces, as in
 the ADM approach  to General Relativity (see e.g. \cite{Misner:1974qy}).
As a result, one finds that Ricci-scalar derivative contributions to
 the evolution equations for   $  h^{(S)} $  are not sufficient for propagating this field: its dynamics is in fact
 constrained to live on the hypersurfaces orthogonal to $v^\mu$, with no components on the direction
of the system evolution.
  Indeed, $  h^{(S)} $ plays a  role analogous to the lapse function ${ N}$ in the ADM formalism. This can also be
  deduced by the definition of  $  h^{(S)} $ in eq \eqref{defcom123}:  this quantity collects the contribution to $h_{\mu\nu}$
  from the components along the vector  $v^\mu$, exactly as the lapse constraint in ADM.
   (We
  also  discuss in Appendix \ref{exafphr}  an explicit, simple example where   $  h^{(S)} $  is manifestly non-dynamical.)

Can the energy-momentum tensor in the right-hand-side of eq \eqref{TildeEFE}, or the Ricci tensor in its left-hand-side  qualitatively change these considerations? Not if it is derived from a covariant action as \eqref{genac1}, where non-minimal couplings
of dark energy scalar to the metric are  expressed in a covariant form in terms of the metric, Riemann, and Ricci tensors. This is always possible  in manifestly  covariant and diffeomorphism invariant  formulations of scalar-tensor systems.
 They   contribute to the scalar kinetic terms with a structure as  the one  we
obtained from the Ricci scalar \eqref{ricS1}: the second derivatives
acting on  $  h^{(S)} $ appear in combinations weighted by the projector $\Lambda_{\mu\nu}$ as  the parenthesis in the last term of eq \eqref{ricS1}. Hence the same considerations as above apply. Since for such space-time slicing  it is a constrained field,  $  h^{(S)} $  does not propagate, and its role is to impose
conditions on the slow-frequency part of the system, or on the remaining high-frequency modes.

\smallskip

Let us then assume to solve the equations of motion for the non-dynamical field  $  h^{(S)} $, and substitute its solution on the original action.
We are left with an action containing
$ h^{(TT)}_{\mu\nu}$  and $ h^{{\rm (tr)}} $
as potentially propagating high-frequency degrees of freedom.
 Thanks
to linearity, the linearized  Einstein equations for the propagating modes can be decomposed as

 \be
G_{\mu\nu}^{(1)} \left[
  h_{\rho\sigma}^{(TT)}
\right]+
G_{\mu\nu}^{(1)} \left[
  h^{{\rm (tr)}}
\right]
\,=\,
T_{\mu\nu}^{(T)} \left[
  h_{\rho\sigma}^{(TT)}
 \right]
 +
T_{\mu\nu}^{\rm{(tr)}} \left[
  h^{{\rm (tr)}}
\right]\,,
\label{sum2eq}
\ee

\bigskip
\noindent
 We expect that second derivatives contributions on the scalar sector have a rich structure, with
 different coefficients in front of contributions orthogonal or parallel to the vector $v^\mu$, associated
 with spontaneous breaking of Lorentz invariance associated with the vector $v^\mu$. As a consequence, tensor
 and scalar fluctuations normally propagate with different velocities \footnote{Recently, various scenarios
   have been analyzed \cite{Garoffolo:2019mna,Garoffolo:2020vtd,Dalang:2020eaj,Ezquiaga:2020dao}
 where, at the price of tunings, tensor and scalar propagate with the same speed. We
 do not consider this case in this work.}.
   Given the strong experimental bounds on the GW velocity associated with the GW170817 event,   we set the speed
 of GW to the one of light.

Within this hypothesis, we 
  implement a geometric optic Ansatz to both the  $ h^{(TT)}_{\mu\nu}$  and $ h^{{\rm (tr)}} $ sectors, using
  the  geometric optic  approach explained in the main text.
 We write:
\bea
\label{ans1t}
 h^{(TT)}_{\mu\nu}&=&{\cal A}_{\mu\nu}^{(T)}\,\exp{\left[
i\,{\psi^{(TT)}}/{\epsilon}
\right]
}\,,
\\
\label{ans1s}
 h^{{\rm (tr)}}&=&{\cal A}^{{\rm (tr)}}\,\exp{\left[
i\,{\psi^{{\rm (tr)}}}/{\epsilon}
\right]}\,.
\eea
The amplitudes of both modes are slowly varying, while the phases are  rapidly varying thanks to the factors of $1/\epsilon$ in the exponent.
 When plugging Ansatz \eqref{ans1t} and \eqref{ans1s} into eq \eqref{sum2eq}, one gets
a linear combination of  terms with  rapidly oscillating phases and slowly varying overall coefficients.
Schematically, we expect that the geometric optics limit of Einstein equations has a structure as
\be
\label{condph12}
\left( \dots\edu \right)\,\exp{\left[
i\,{\psi^{(TT)}}/{\epsilon}
\right]
}+\left( \dots\edu \right)\,\exp{\left[
i\,{\psi^{{\rm (tr)}}}/{\epsilon}
\right]
}\,=\,0
\ee
where within the parenthesis we collect slowly varying contributions at order $\epsilon^{-2}$ and $\epsilon^{-1}$
in a gradient expansion. The $\epsilon^{-2}$ contributions  depend on derivative of the phases $\psi^{(TT)}$ and  $\psi^{{\rm (tr)}}$: they
control the dispersion relations for the two species  of excitations, scalar  and GW (see section \ref{secGeoAN}
for a geometric optics analysis of the GW sector).
  Since in general  $ h^{(TT)}_{\mu\nu}$  and $ h^{{\rm (tr)}} $
propagate with different speed, they are characterized by distinct dispersion relations, hence the phases
$\psi^{(TT)}$ and  $\psi^{{\rm (tr)}}$ are different.
   Equation \eqref{condph12} is a linear combination of two contributions weighted by  two distinct phases which rapidly oscillate
  over space and time:
 in
   order to satisfy it,
       we need
to impose that the  coefficients of each of these two terms  separately vanish.
Within the geometric optics limit,
this procedure effectively separates the evolution of scalar
modes (characterized by the phase $\psi^{{\rm (tr)}}$) and GW modes (characterized by the phase $\psi^{(TT)}$).

\smallskip

 Given these considerations,  in dark energy scenarios where the scalar
 and GW
 modes have
 different phases due to different dispersion relations,
 we
 can effectively {\it separate} the high-frequency  GW and scalar sectors in eq \eqref{sum2eq}, and write
 \bea
\label{genStS}
G_{\mu\nu}^{(1)}  \left[
h^{{\rm (tr)}}
\right]
&=&
T_{\mu\nu}^{{\rm (tr)}} \left[
h^{{\rm (tr)}}\right]
\,,
\\
\label{genStT}
G_{\mu\nu}^{(1)} \left[
 h_{\rho\sigma}^{(TT)}
\right]
&=&
T_{\mu\nu}^{(T)} \left[
 h_{\rho\sigma}^{(TT)}\right]\,.
\eea
  Within our hypothesis, GW sector is decoupled from the scalar sector at
the linearized level, and we can study its dynamics as done in the main text. The arguments discussed above might be made   more rigorous with a more systematic and detailed analysis
 of perturbations evolution equations, for example using the approach of the recent work \cite{Fier:2021fbt}.  We leave this analysis to separate investigations.

\section{A simple example: $F(\phi)\,R$}\label{exafphr}

Let us make  a specific, simple example of the friction-term contributions found in our general formula
of eq \eqref{singpa1}, which arises in models characterized by a time-varying Planck mass controlled
by the dark energy scalar field $\phi$.  We consider the following non-minimal kinetic coupling between scalar $\phi$ and metric
\be
\label{FRcou1}
{\cal L}\,=\,F(\phi) R\,,
\ee
which can be considered a part of the classic Brans-Dicke action \cite{Brans:1961sx}.  We are not interested
to study in detail the system, but only apply to its corresponding GW  evolution equations the approach explained in Appendix \ref{sec_theory1}.

We decomponse the linearized
Einstein equations in terms of the high-energy fluctuations,
and focus on orders $1/\epsilon^2$ and  $1/\epsilon$ in a gradient expansion,
as described in the previous section \ref{sec_theory1}. We find that GW modes obey the equation

\be
 \label{BDevGW}
\Box  h^{(TT)}_{\mu\nu}\,=\,\frac{2 F_{, \phi}}{F}\,v^\lambda\,\nabla_\lambda  h^{(TT)}_{\mu\nu}\,.
\ee
An evolution equation governing scalar modes can be determined by taking the trace of the Einstein equations
\bea
&&
 \Box h^{\rm (tr)}- \Lambda_{\alpha \beta}\,\nabla^\alpha \nabla^\beta
\left(h^{(S)}+\frac13 h^{\rm (tr)}\right)
\,=\,-\,\frac{3\,\sqrt{2 X} F_{,\phi}}{F}\, X^\lambda \nabla_\lambda  \left(h^{(S)}
+h^{\rm (tr)}\right)\,,
\label{ex2eqs}
\eea
where the vector $X_\mu$ is defined in eq \eqref{defXmu1}, and the projector $\Lambda_{\mu\nu}$ in eq \eqref{projector}.

 These equations have  the  structure
expected from our considerations in the main text and
in Appendix \ref{sec_theory1}. In  fact, comparing the GW evolution equation \eqref{BDevGW}
with the general expression in eq \eqref{singpa1}, we notice that the former
has a friction term ${\cal T}$
 controlled by the derivative of $F$ along the dark energy field: ${\cal T}\,=\,{2 F_{, \phi}}/{F}$.  Using
the results of section \ref{sec-GWEMT},  we find that the energy-momentum-tensor at second order
in the transverse-traceless fluctuations reads (we choose the extreme of integration $ \phi_{\rm in}$ such that $F( \phi_{\rm in})=1$)
\bea
T_{\mu\nu}^{\rm (2),\,ST}&=&\epsilon^{2}\,\frac{e^{-\int {\cal T}}}{32\,\pi}\,\langle \nabla_{\mu}  h^{(TT)}_{\rho\sigma}
\,\nabla_{\nu}  h^{(TT)\,\,\rho\sigma}\rangle
\nonumber
\\
&=&\epsilon^{2}\,\frac{e^{\int_{ \phi_{\rm in}}^\phi \frac{d \ln{F}}{d \tilde \phi} \,d \tilde \phi}}{32\,\pi}\,\langle \nabla_{\mu}  h^{(TT)}_{\rho\sigma}
\,\nabla_{\nu}  h^{(TT)\,\,\rho\sigma}\rangle
\nonumber
\\
&=&\epsilon^{2}\,\frac{F(\phi)}{32\,\pi}\,\langle \nabla_{\mu}  h^{(TT)}_{\rho\sigma}
\,\nabla_{\nu}  h^{(TT)\,\,\rho\sigma}\rangle\,,
\eea
which is  the expected structure associated with the Lagrangian of eq \eqref{FRcou1}.
 The scalar modes
  have a kinetic structure depending on the Lorentz violating vector $v^\mu$. The kinetic term
  for $h^{(S)}$ in eq   \eqref{ex2eqs}  is projected by the tensor  $\Lambda_{\mu\nu}$ in
  the direction orthogonal to $v^\mu$, as discussed in Appendix \ref{sec_theory1}, hence
  it does not contribute to the dynamical degrees of freedom.

\smallskip

We can apply these findings to cosmology, and consider the case of GW propagating through a conformally
flat Friedmann-Robertson-Walker (FRW) universe, with metric $d s^2\,=\,a^2(\eta)\,\eta_{\mu\nu}\,d x^\mu d x^\nu$, and for a homogeneous scalar field $\bar \phi\,=\,\bar \phi(\eta)$. Eq \eqref{BDevGW} results then  (${\cal H}\,=\,a'/a$, with prime denoting derivative along time)
\be
h^{(TT)''}_{\mu\nu}+2  \,{\cal H}\,\left( 1-\frac{F_{,\phi}}{F}\,\frac{\bar \phi'}{\cal H}\right)\,h^{(TT)'}_{\mu\nu}
-{\bf \nabla}^2 h^{(TT)}_{\mu\nu}\,=\,0\,.
\ee
The effect of the friction term due to the non-minimal scalar-tensor couplings has the expected structure and is  manifest within the parenthesis of the
previous expression.

It would be interesting to further apply our covariant approach to fluctuations to more
complex scenarios, from Horndeski \cite{Horndeski:1974wa} to beyond Horndeski \cite{Gleyzes:2014dya}
and DHOST \cite{Langlois:2015cwa,Crisostomi:2016czh,BenAchour:2016fzp}.

\section{Gauge invariance of the GW energy-momentum tensor}\label{appA}

We show that the structure of eq
\eqref{tmngw2} is fixed by the gauge invariance. The most general structure
for the energy momentum tensor quadratic in the high-frequency transverse-traceless (TT) modes $ h^{(TT)}_{\mu\nu}$
is
\be\label{tmngw2b}
T_{\mu\nu}^{\rm (2),\,MG}\,=\,\epsilon^{2}\,\frac{1}{32\,\pi}\,\langle \nabla_{\mu}  h^{(TT)}_{\alpha \beta}
\,\nabla_{\nu}  h^{(TT)}_{\gamma \delta}\rangle\,{\cal C}^{\alpha \beta \gamma \delta}\,,
\ee
where ${\cal C}^{\alpha \beta \gamma \delta}$ depends on slowly-varying fields. This is the most
general structure for $T_{\mu\nu}^{\rm (2),\,MG}$ compatible with the   TT gauge
imposed on $ h^{(TT)}_{\mu\nu}$.
After fixing the TT gauge as discussed in section \ref{sec_theory1}, we are left with invariance under the transformation
 in eq \eqref{finressA}, as derived around eq \eqref{resTTtr1}.
 We need to ensure that $T_{\mu\nu}^{\rm (2),\,MG}$ is invariant as well.
Notice that the tensor ${\cal C}^{\alpha \beta \gamma \delta}$
has to be invariant under an interchange of $\alpha$ and $\beta$.
 Under such  gauge transformation, at the linearized level we get
a contribution
\bea
\delta T_{\mu\nu}^{\rm (2),\,MG}&=&
-\epsilon^{2}\,\frac{1}{16\,\pi}\,\langle \nabla_{\mu} \nabla_{\alpha} \xi^{(T)}_{\beta}
\,\nabla_{\nu}  h^{(TT)}_{\gamma \delta}\rangle\,{\cal C}^{\alpha \beta \gamma \delta}
\,,
\\
&=&
\epsilon^{2}\,\frac{{ 1}}{16\,\pi}\,\langle \nabla_{\mu} \xi_{\beta}^{(T)}
\,\nabla_{\nu}  \nabla_{\alpha}  h^{(TT)}_{\gamma \delta}\rangle\,{\cal C}^{\alpha \beta \gamma \delta}
\,,
\eea
which must vanish for any $ \xi_{\beta}^{(T)}$, and for any choice of $\mu$, $\nu$. The only way to ensure this is
to use the transverse gauge condition, and require that the contraction with ${\cal C}^{\alpha \beta \gamma \delta}$ forces
 the condition $\alpha=\gamma$ (or alternatively $\alpha=\delta$). This implies
\be
{\cal C}^{\alpha \beta \gamma \delta}
\,=\,C\,\delta^{\alpha\gamma} \,C^{\beta \delta}
\ee
for some constant $C$. Plugging this  result in eq \eqref{tmngw2b} we get
\be\label{tmngw2bB}
T_{\mu\nu}^{\rm (2),\,MG}\,=\,\epsilon^{2}\,\frac{C}{32\,\pi}\,\langle \nabla_{\mu}  h^{(TT),\,\alpha}_{\,\, \beta}
\,\nabla_{\nu}  h^{(TT)}_{\alpha \delta}\rangle\,{\cal C}^{ \beta  \delta}\,.
\ee
Since the EMT is symmetric in the indexes, the  quantity ${\cal C}^{ \beta  \delta}$ is symmetric. Applying
the transformation
\eqref{finressA}, we find the only non-vanishing contribution
\bea
\delta T_{\mu\nu}^{\rm (2),\,MG}&=&
-\epsilon^{2}\,\frac{C}{16\,\pi}\,\langle \nabla_{\mu} \nabla_{\beta} \xi^{(T)}_{\alpha}
\,\nabla_{\nu}  h^{(TT),\,\alpha}_{\,\, \delta}\rangle\,{\cal C}^{ \beta  \delta}\,,
\\
&=&
\epsilon^{2}\,\frac{{\cal C}}{16\,\pi}\,\langle \nabla_{\mu} \xi^{(T)}_{\alpha}
\,\nabla_{\nu}  \nabla_{\beta}  h^{(TT),\,\alpha}_{\,\, \delta}\rangle\,{\cal C}^{ \beta  \delta}\,,
\eea
and
the only way to make it always vanishing is to have ${\cal C}^{ \beta  \delta}\propto \delta^{\beta \delta}$. Symmetry
 arguments force the EMT to be proportional to Isaacson's form \eqref{tmngw2}, up to the overall constant, as
we wish to prove.

\section{Comparison with the literature}
\label{app-compLISA}

We now that our expression \eqref{ratiolu1} for the ratio among GW and electromagnetic luminosity distances:
\be \label{ratiolu1a}
\frac{d_L^{\rm (GW)}}{d_L^{\rm (EM)}}\,=\,
\exp{\left[-\frac12\,\int^{\lambda_S}_{0}\,{\cal T}\,\frac{d \bar \phi}{d \lambda}\,d \lambda \right]}
\,,
\ee
coincides
with analog expressions found in the literature, once we specialize to a cosmological setting.

Let us then analyze GW travelling through an unperturbed,
 conformally flat Friedmann-Robertson-Walker (FRW) universe with
metric $d s^2\,=\,a^2(\eta)\left( -\d \eta^2+d \vec x^2\right)$. The dark-energy background scalar is time-dependent only, $\bar\phi\,=\,\phi(\eta)$, and we have $v_\mu\,=\,( \bar \phi',\,0,\,0,\,0)$.
Let us use, for definiteness, the notation of \cite{Belgacem:2019pkk}. We understand tensorial indexes and call
 $h^{(TT)}_{\mu\nu}\,=\,h(\eta, \vec x)$.
 The evolution equation for the GW modes is expressed as (${\cal H}\,=\,a'/a$)
\be \label{eqplisa}
h''+2 {\cal H}\,\left(1-\delta (\eta) \right)\,h'-\nabla^2 h\,=\,0\,,
\ee
where  modified gravity friction contributions are contained in the time-dependent
parameter $\delta (\eta)$. The expression for the ratio between
GW and electromagnetic luminosity distances is found to be (see  \cite{,Belgacem:2019pkk})
\bea \label{lisratio}
\frac{d_L^{\rm (GW)}}{d_L^{\rm (EM)}}&=&\exp{\left[-\,\int_0^z\,\frac{\delta (z')}{1+z'}\,d z'\right]}
\,.
\eea

\medskip
 We now apply our formula \eqref{ratiolu1} to this cosmological example. Equation \eqref{singpa1} reads
   \be
h''+2 {\cal H}\,\left(1-\frac{{\cal T}\,\bar \phi'}{2 {\cal H}}\right)\,h'-\nabla^2 h\,=\,0
\,.
\ee
Comparing with eq \eqref{eqplisa}, we can then identify $\delta\,=\,{{\cal T}\,\bar \phi'}/{(2 {\cal H})}$.
Using also the fact that $d \eta/d t\,=\,1/a$,  ${\cal H}\,=\,H/a$, $1+z\,=\,a(0)/a(t)$, we can write eq \eqref{lisratio}
as (the suffix $0$ indicate present-day observers)
\bea
\frac{d_L^{\rm (GW)}}{d_L^{\rm (EM)}}&=&
\exp{\left[-\,\int_0^{z_s}\,\frac{\delta (z)}{1+z}\,d z\right]}
\,=\, \exp{\left[-\,\int^{t_s}_{t_0}\,{\delta (t)}\,H\,d t\right]} \,=\, \exp{\left[-\,\int^{\eta_s}_{\eta_0}\,{\delta (\eta)}\,{\cal H}\,d \eta\right]}
\,,
\nonumber
\\
&=&\exp{\left[-\,\int^{\eta_s}_{\eta_0}\,\frac{{\cal T}\,\bar \phi'}{2 {\cal H}}\,{\cal H}\,d \eta\right]}
\,=\,
\exp{\left[-\,\frac12\,\int^{\eta_s}_{\eta_0}\,{{\cal T}\,\bar \phi'}\,d \eta\right]}
\,,
\nonumber
\\
&=&
\exp{\left[-\,\frac12\,\int^{\lambda_s}_0\,{{\cal T}}\,\frac{d  \bar \phi}{d \lambda} d \lambda\right]}\,,
\eea
which coincides with eq \eqref{ratiolu1a}.

\section{Proof of the step between eq \eqref{forAp1} and eq \eqref{ether1} }\label{appEthP}

We prove the validity of the   step between eq \eqref{forAp1} and eq \eqref{ether1}, using results from \cite{Sasaki:1987ad}. For shortening the notation, we call
\be
q(\lambda)\,=\,\delta(R_{\mu\nu} K^\mu K^\nu )_\lambda
\ee
the geometrical combination depending on the background geometry, which implicitly appears in both formulas
 \eqref{forAp1} and eq \eqref{ether1}.

 We call
 \bea
I_A&=&\int_0^{\lambda_s} \delta \theta_A\,d \lambda\,=\,-\int_0^{\lambda_s} \frac{d\lambda}{\lambda^2}
\int_0^\lambda\,d \lambda'\,\lambda'^2\,q(\lambda')\,,
\\
I_L&=&\int_0^{\lambda_s} \delta \theta_L\,d \lambda\,=\,\int_0^{\lambda_s} \frac{d\lambda}{(\lambda-\lambda_s)^2}
\int_\lambda^{\lambda_s}\,d \lambda'\,(\lambda'-\lambda_s)^2\,q(\lambda')\,.
\eea
We wish to prove
\be
I_L+I_A\,=\,0\,.
\ee
First we re-express $I_L$ more conveniently. We change variable $\lambda=\lambda_s-\sigma$, $\lambda'=\lambda_s-\sigma'$. Then
\be
I_L\,=\,
\int_0^{\lambda_s} \frac{d\sigma}{\sigma^2}
\int_{0}^{\sigma}\,d \sigma'\,\sigma'^2\,q(\lambda_s-\sigma')
\ee
 Then by an integration by parts we can re-write  $I_A$  as a single integral as
\bea
I_A&=&\int_0^{\lambda_s} d \lambda \left(\frac{\lambda^2}{\lambda_S}
-\lambda \right)\,q(\lambda)\,.
\eea
Moreover, simple steps lead to
\bea
I_L&=&-
\int_0^{\lambda_s} d \lambda \left(\frac{\lambda^2}{\lambda_s}
-\lambda \right)\,q(\lambda_s-\lambda)\,,
\\
&=&-\int_0^{\lambda_s} d \lambda \left[\frac{(\lambda_s-\lambda)^2}{\lambda_s}
-(\lambda_s-\lambda) \right]\,q(\lambda)\,,
\\
&=&-I_A\,,
\eea
as desired.

\section{The geometric time-delay}
\label{app-timedelay}

 \begin{figure}[h!]
\centering
 \includegraphics[width = 0.42 \textwidth]{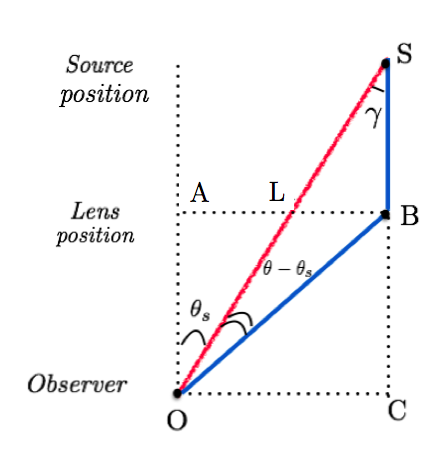}
 \caption{\it The configuration we consider.}
 \label{fig:cons}
\end{figure}

We derive the expression for the geometric time delay of waves whose trajectory is bended
by a point-like lens, in the limit of geometric optics and of Euclidean geometry. The
GW moves with the speed of light,  and we  consider Fig \ref{fig:cons}  are reference.
 In Fig \ref{fig:cons}
   the angle $\theta-\theta_s$  represented the difference
between the lensed and unlensed angular quantities.  We denote with $\ell$ the length of the lines: for
example $\ell_{AL}$ is the length of the line that joins point $A$ with point $L$. Angular distances
are defined  as ratios between lengths and angles they subtend with respect to
who observes them. For example
\bea
 D_{OL}&=&\ell_{AL}/\theta_s \hskip1cm,\hskip1cm D_{SL}=\ell_{LB}/\gamma
  \hskip1cm,\hskip1cm
D_{SO}\,=\,{\ell_{OC}}/{\gamma}\hskip1cm,\hskip1cm
\eea
(we should take
care at the position of the indexes in the angular distances). We work in the limit of infinitesimal angles, so we can expand
trigonometric functions. Hence
\be
\ell_{OL} \,\sin \theta_s\,=\,\ell_{AL}\hskip1cm\Rightarrow\hskip1cm \ell_{OL} \,=\,D_{OL}\,.
\ee
Similarly, one has $\ell_{OB}\,=\,D_{OL}$.
 We compute step by step  the GW time delay,  corresponding to the quantity (recall that GW travel at the
 speed of light, set to one)
\be
\Delta t\,=\,\ell_{SB}+\ell_{OB}-\ell_{SO}\,.
\ee

Since the triangles $LSB$ and $OSB$ are similar, we can write the equality
\be
\frac{\ell_{LS}}{\ell_{OS}}\,=\,\frac{\ell_{LB}}{\ell_{OC}}\,=\,\frac{D_{SL}}{D_{SO}}\,.
\ee
Hence
\be
\ell_{OS}\,=\,\ell_{OL}+\ell_{LS}\,=\,\ell_{OL}+\ell_{OS}\,\frac{D_{SL}}{D_{SO}}\,.
\ee
This fact implies that
\be
\ell_{OS}\,=\,D_{OL}\, \left(1-\frac{D_{SL}}{D_{SO}} \right)^{-1}\,=\,\frac{D_{OL} D_{SO}}{D_{SO}-D_{SL}}\,.
\ee

Moreover, the law of cosines ensures that
\be
\ell_{SB}^2\,=\,\ell_{OB}^2+\ell_{OS}^2-2\,\ell_{OB} \ell_{OS}\,\cos(\theta-\theta_s)\,.
\ee
Expanding the cosine for small angles, we can reassemble the previous formula as
\bea
\ell_{SB}&\simeq&\left( \ell_{OS}-\ell_{OB} \right) \sqrt{1+\frac{\ell_{OB} \ell_{OS}}{(\ell_{OB}-\ell_{OS})^2} |\theta-\theta_s|^2 }\,,
\\
&\simeq& \left( \ell_{OS}-\ell_{OB} \right) \,\left(1+ \frac{\ell_{OB} \ell_{OS}}{2(\ell_{OB}-\ell_{OS})^2} |\theta-\theta_s|^2 \right)\,.
\eea
Then the quantity we are after is
\bea
\Delta t&=&\frac{\ell_{OB} \ell_{OS}}{2(\ell_{OS}-\ell_{OB})} |\theta-\theta_s|^2\,=\,
\frac{D_{OL}}{2}\,\frac{D_{OL} D_{SO}}{D_{SO}-D_{SL}}\,\frac{1}{\frac{D_{OL} D_{SO}}{D_{SO}-D_{SL}}-D_{OL}}\,|\theta-\theta_s|^2\,,
\\
&=&\frac{D_{OL} D_{SO}}{2 \,D_{SL}}\,|\theta-\theta_s|^2\,,
\eea
which is the formula used in eq \eqref{genexTD} of the main text.

\end{appendix}

\addcontentsline{toc}{section}{References}
\bibliographystyle{utphys}

\mciteSetMidEndSepPunct{}{\ifmciteBstWouldAddEndPunct.\else\fi}{\relax}

\providecommand{\href}[2]{#2}\begingroup\raggedright\endgroup

\end{document}